\documentclass[12pt,letterpaper]{article}

\usepackage{fullpage}
\usepackage{Alegreya}
\usepackage{hyperref}
\usepackage[compatibility=false]{caption}
\usepackage{subcaption}
\usepackage{amsmath}
\usepackage{graphicx}
\usepackage{natbib}
\usepackage{comment}
\usepackage{booktabs}
\usepackage{booktabs}
\usepackage{multirow}
\usepackage[framemethod=TikZ]{mdframed}

\usepackage{tikz}
\usetikzlibrary{bayesnet}
\usepackage{color,colortbl}

\mdfdefinestyle{alert}{%
    linecolor=red,
    outerlinewidth=3pt,
    roundcorner=10pt,
    innertopmargin=10pt,
    innerbottommargin=10pt,
    innerrightmargin=10pt,
    innerleftmargin=10pt,
    backgroundcolor=white}

\begin{document}

\title{Exploring Author Gender in Book Rating and Recommendation}


\author{Michael D. Ekstrand and Daniel Kluver}



\date{Aug. 19, 2019}

\maketitle

\begin{mdframed}[style=alert]
  \textbf{Notice:} this is an unreviewed preprint of an extended version of our RecSys paper \citep{Ekstrand2018-ip}.
  It has been submitted for journal publication; before building on or citing this work, see \url{https://md.ekstrandom.net/pubs/bag-extended} for updates
  or the arXiv abstract page for the published DOI.
  \textbf{Do not cite this version after October 1, 2020.}
\end{mdframed}

\begin{abstract}
  Collaborative filtering algorithms find useful patterns in rating and consumption data and exploit these patterns to guide users to good items. Many of the patterns in rating datasets reflect important real-world differences between the various users and items in the data; other patterns may be irrelevant or possibly undesirable for social or ethical reasons, particularly if they reflect undesired discrimination, such as discrimination in publishing or purchasing against authors who are women or ethnic minorities. In this work, we examine the response of collaborative filtering recommender algorithms to the distribution of their input data with respect to a dimension of social concern, namely content creator gender. Using publicly-available book ratings data, we measure the distribution of the genders of the authors of books in user rating profiles and recommendation lists produced from this data. We find that common collaborative filtering algorithms differ in the gender distribution of their recommendation lists, and in the relationship of that output distribution to user profile distribution.
\end{abstract}

\section{Introduction}

The evaluation of recommender systems has historically focused on the accuracy of recommendations \citep{Herlocker2004-rh,Shani2010-wj}.  
When it is concerned with other characteristics, such as diversity, novelty, and user satisfaction \citep{Hurley2011-cg,Ziegler2005-zo,Knijnenburg2012-ia}, it often continues to focus on traditionally-understood information needs.  But this paradigm, while irreplaceable in creating products that deliver immediate value, does not tell the whole story of a recommender system's interaction with its users, content creators, and other stakeholders.

In recent years, public and scholarly discourse has subjected artificial intelligence systems to increased scrutiny for their impact on their users and society. Much of this has focused on classification systems in areas of legal concern for discrimination, such as criminal justice, employment, and housing credit decisions. However, there has been interest in the ways in which more consumer-focused systems such as Uber \citep{Rosenblat2016-kb}, TaskRabbit \citep{Hannak2016-wy}, and search engines \citep{Magno2016-qi} interact with issues of bias, discrimination, and stereotyping.

Social impact is not a new concern in recommender systems.  \emph{Balkanization} \citep{Van_Alstyne2005-vn} or \emph{filter bubbles}, popularized by \citet{Pariser2011-wu}, are one example of this concern: do recommender systems enrich our lives and participation in society or isolate us in echo chambers?
Recommender systems are intended to influence their users' behavior in some way; if they did not, there would be little reason to operate them. Understanding the ways in which recommender systems actually interact with past, present, and future user behavior is a prerequisite to assessing the ethical, legal, moral, and social ramifications of that influence.

In this paper, we present experimental strategies and observational results from our investigation into how recommender systems interact with author gender in book data and associated consumption and rating patterns.
The direct experimental outcomes of this paper characterize the distribution of author genders in existing book data sets and the response of widely-used collaborative filtering algorithms to that distribution, and assess the accuracy impact of deploying efficient strategies for adjusting the gender makeup of recommendation lists.
The data and methods that we have used for this paper, however, extend beyond our immediate questions and we expect them to be useful for much more research on fairness and social impacts of recommender systems.
Our data processing, experiments, and analysis are all reproducible from public data sets with the code accompanying this paper.

Our experiments address the following questions:

\begin{description}
\item[RQ1] How are author genders distributed in book catalog data?
\item[RQ2] How are author genders distributed in users' book reading histories?
\item[RQ3] What is the distribution of author genders in the recommendations users receive from common collaborative filtering algorithms? This measures the \emph{overall} behavior of the recommender algorithm(s) with respect to author gender.
\item[RQ4] How do individual users’ gender distributions propagate into the recommendations that they receive? This measures the \emph{personalized} gender behavior of the algorithms.
\item[RQ5] What is the loss in accuracy when adjusting the distribution of recommendation lists to achieve a particular target?
\end{description}

While we expect recommender algorithms to propagate patterns in their input data, due to the general principle of ``garbage in, garbage out'', the particular ways in which those patterns do or do not propagate through the recommender is an open question.
Recommender systems do not always propagate all input data patterns~\citep{Channamsetty2017-ri}, and it is important to understand how this (non-)propagation relates to matters of social concern.

\subsection{Motivation and Fairness Construct}

The work in this paper is motivated by our concern for issues of \emph{representation} in book authorship.
There are efforts in many segments of the publishing industry to improve representation of women, ethnic minorities, and other historically underrepresented groups.
Multiple organizations undertake counts of books and book reviews to assess the representation of women and nonbinary individuals in the literary landscape~\citep{Pajovic2016-bb,Vida2017-bn}.

Our goal is to understand how recommendation algorithms interact with such efforts.
When a new author gets their first book published and it appears on recommender-driven platforms, whether they are sales channels like Amazon or Barnes and Noble or reading communities like GoodReads, does the recommender system help them their work the audience that will propel them to success?  Is it a neutral path, neither helping nor hindering?  Or is algorithmic recommendation another hurdle to their success, stacking the deck in favor of well-known authors and the status quo of the publishing industry?

Author representation also has a consumer-facing dimension: what picture does a book service's discovery layer paint of the space of book authorship?
When a user is looking for books, do they see books by a diverse range of authors, or are the books that are surfaced focused on certain corners of the authorship space?
This is admittedly a complex question, because recommending books that are not relevant to a user's interests or information need just because of their author's demographics does not make for an effective recommendation or information retrieval system.
Fairness in recommendation needs to be understood in the context of accuracy and other measures of effectiveness.

The experiments in this paper are focused on \emph{consumer-centered provider fairness}.
Our framing is similar to ``calibrated fairness'' proposed by \citet{steck_calibrated_2018}, in that we are concerned with the makeup of recommendation lists and their connection to users' input profiles.
While there are many ways of conceiving of provider fairness, some of which we examine in Section~\ref{sec:fia}, list composition seems particularly well-suited to understanding representation as it is experienced by users of the system.

\subsection{Contributions and Summary of Findings}

\textbf{Data Integration.}
We describe an integration of six different public data sets to study social issues in book recommendation, cataloging and justifying the data linking decisions we made along the way.
This composite data set itself will be useful for further research, and work looking at other domains and data sources will need to make similar sets of decisions.
Our strategy, therefore, serves as a case study in obtaining and preparing data for fairness and social impact research in recommender systems and likely in other domains as well.

\textbf{Experimental Methodology.}
We describe an end-to-end experimental pipeline and statistical analysis for studying representation and list composition in recommendation, and how user patterns do or do not propagate into recommendation outputs.

\textbf{RQ1--2: Gender in Users and Catalogs.}
We find that book rating data has a higher proportion of female-authored books, across multiple rating data sets from different sites, than the underlying book corpus as estimated through the Library of Congress.
GoodReads is close to gender parity in its user-book interactions.
Individual user tendencies are highly diffuse, with an average tendancy that is in line with the set of rated books.

\textbf{RQ3--4: Recommender Response.}
We find that several collaborative filtering algorithms propagate much of the user's input tendancy with respect to gender into their output: users who rate more female-authored books receive more female-authored recommendations.

\textbf{RQ5: Loss for Targeting Distributions.}
We find that arbitrary recommendation list distributions, such as equal balance between male and female authors, can be achieved with little loss in recommendation accuracy.

The purpose of this paper is not to make any normative claims regarding the distributions we observe, simply to describe the current state of the data and algorithms.
We do not currently have sufficient data to determine whether the distributions observed in available data represents under- or over-representation, or what the ``true'' values are.
We hope that our observations can be combined with additional information from other disciplines and from future work in this space to develop a clearer picture of the ways in which recommender systems interact with their surrounding sociotechnical ecosystems.
Our normative claim is that researchers and practitioners should care and seek to understand how their systems interact with these issues, and our methods provide a starting point for such experiments.
\section{Background and Related Work}

Our present work builds on work in both recommender systems and in bias and fairness in algorithmic systems more generally.

\subsection{Recommender Systems}

Recommender systems have long been deployed for helping users identify relevant items amongst large sets of possibilities \citep{Ekstrand2010-wg,Adomavicius2005-jr}.
Of particular interest to our current work is \emph{collaborative filtering} (CF) systems, which use patterns in user-item interaction data to estimate which items a particular user is likely to find useful.

While recommender evaluation and analysis often focuses on the accuracy of recommendations \citep{Herlocker2004-rh,Shani2010-wj}, there has been significant work on non-accuracy dimensions of recommender behavior. Perhaps the best-known is diversity \citep{Ziegler2005-zo}, sometimes considered along with novelty \citep{Hurley2011-cg,Vargas2011-fq}.  \citet{Lathia2010-cr} examined the \emph{temporal} diversity of recommender systems, studying whether they changed their recommendations over time.  Other work has quantified recommendation bias with respect to classes of items \citep{Jannach2015-yf}.

\subsection{Social Impact of Recommendations}

Recommender systems researchers have been concerned for how recommenders interact with various individual and social human dynamics.  One example is balkanization or filter bubbles \citep{Van_Alstyne2005-vn,Pariser2011-wu}, mentioned earlier; recent work has sought to detect and quantify the extent to which recommender algorithms create or break down their users' information bubbles \citep{Nguyen2014-se} and studied the effects of recommender feedback loops on users' interaction with items \citep{Hosanagar2013-dp}.

Other work seeks to use recommender technology to promote socially-desirable outcomes such as energy savings \citep{Starke2015-hn}, better encyclopedia content \citep{Cosley2007-tf}, and new kinds of relationships \citep{Resnick2001-hq}.

\subsection{Representation in the Book Industry}
There are efforts in many segments of the publishing industry to improve representation of women, ethnic minorities, and other historically underrepresented groups.
Multiple organizations undertake counts of books and book reviews to assess the representation of women and nonbinary individuals in the literary landscape~\citep{Pajovic2016-bb,Vida2017-bn}.
We seek to understand how recommendation algorithms interact with such efforts: are they a help, a hindrance, or a neutral conduit?

\subsection{Bias and Fairness in Algorithmic Systems}

Questions of bias and fairness in computing systems are not new; \citet{Friedman1996-mn} considered early on the ways in which computer systems can be (unintentially) biased in their design or impact. In the last several years, there has been increasing interest in the ways that machine learning systems are or are not fair. \citet{Dwork2012-ai} and \citet{Friedler2016-pb} have presented definitions of what it means for an algorithm to be \emph{fair}. \citet{Feldman2015-ra} provide a means to evaluate arbitrary machine learning techniques in light of \emph{disparate impact}, a standard for the fairness of decision-making processes adopted by the U.S. legal system.

Bias and discrimination often enter a machine learning system through the input data: the system learns to replicate the biases in its inputs.  This has been demonstrated in word embeddings \citep{Bolukbasi2016-uk} and predictive policing systems \citep{Lum2016-wp,Ensign2017-db}, among others.

\subsection{Fair Information Access}
\label{sec:fia}

\citet{Burke2017-ne} lays out some of the ways in which questions of fairness can apply to recommender systems. In particular, he considers the difference between ``C-fairness'', in which consumers or users of the recommender system are treated fairly, and ``P-fairness'', where the producers of recommended content receive fair treatment.  \citet{Burke2018-bn} and \citet{Yao2017-vz} have presented algorithms for C-fair collaborative filtering, and \citet{Ekstrand2018-ip} examine C-fairness in the accuracy of recommendation lists.

Our present study focuses on P-fairness.  This dimension is somewhat related to historical concerns such as long-tail recommendation and item diversity \citep{Jannach2015-yf}.  \citet{Kamishima2018-ri} and \citet{beutelFairnessRecommendationRanking2019} have presented algorithms for P-fair recommendation; calibration \citep{steck_calibrated_2018} can be viewed as another kind of provider fairness.

\citet{biegaEquityAttentionAmortizing2018} provide metrics for assessing fair \emph{exposure} to providers; this metric assess whether providers are recommended an ``appropriate'' number of times.
Other approaches to assesssing the fairness of rankings look at the makeup of the ranking or prefixes thereof \citep{yangMeasuringFairnessRanked2017, sapiezynskiQuantifyingImpactUser2019}; this is closer to our present work, in which we try to understand how lists are composed from the perspective of gender representation.

In this paper, we present an offline empirical analysis of the calibrated provider fairness of several classical collaborative filtering algorithms and their underlying training data.

\section{Data Sources and Integration}

Traditional recommender systems experiments typically rely on rating or consumption data.
There is a wide range of such data sets publicly available, including movie ratings from MovieLens \citep{harper2016movielens}, product reviews from Amazon \citep{McAuley2015-rq}, and artist play logs from Last.fm \citep{Celma:Springer2010}.
Sometimes these data sets are augmented with additional data, such as additional sources of item data or text crawled from Web pages.
Studying fairness and other social dimensions of recommendation, however, require data that is not commonly provided with rating data \citep{Ekstrand2018-ip}, requiring some creativity.

Investigating how content creator demographics relate to recommendation requires the following classes of data:

\begin{itemize}
    \item \emph{Consumption data} on books users have read and/or rated, to understand reading patterns and train recommendation algorithms.
    \item \emph{Book data} describing books and, for our purposes, their authors.
    \item \emph{Author data} describing the authors themselves, and including demographic characteristics of interest.
\end{itemize}

\begin{figure}
    \centering
    \includegraphics[width=\textwidth]{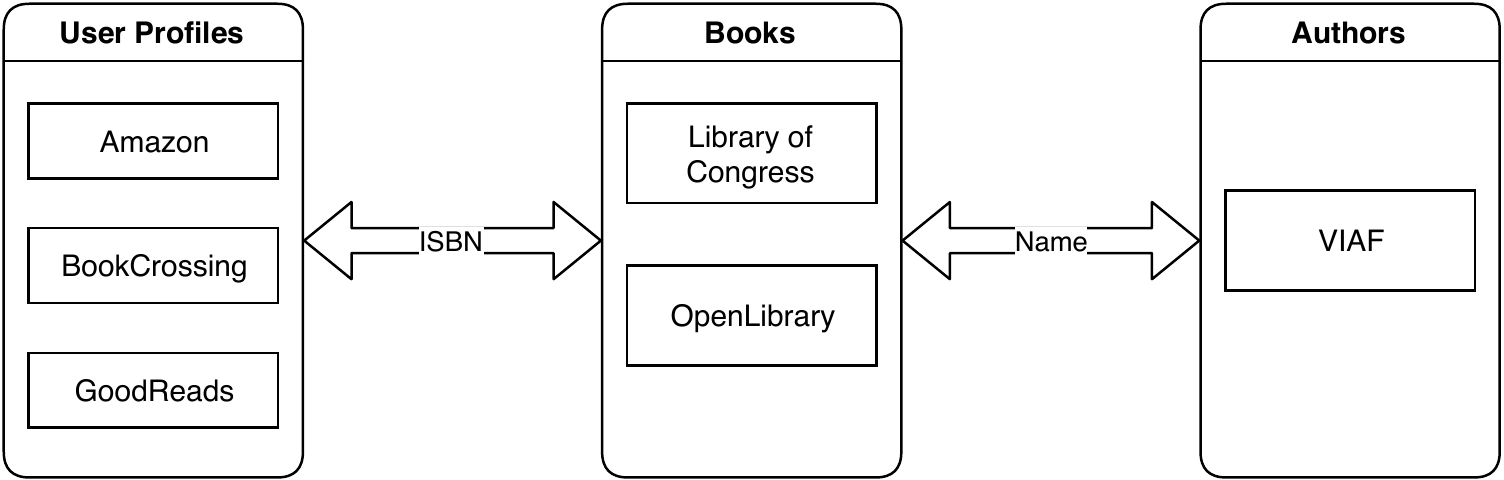}
    \caption{Data set relationships.}
    \label{fig:data-arch}
\end{figure}

Fig.~\ref{fig:data-arch} shows how these data classes fit together and the data sets we use for each.
Linking the data sets together is not easy, due both to the messiness of the data itself (e.g. malformed ISBNs) and the lack of linking identifiers.

This section provides details on our data integration, justifications of data linking decisions we made, and descriptive statistics of the resulting composite data set.

\subsection{Technical Infrastructure}

We use PostgreSQL 11 for storing and integrating the data sources.
The input data comes in variety of formats (including CSV files, JSON objects, and MARC records), and PostgreSQL's flexible data types (including native support for JSON) enable us to express almost all data integration logic as SQL queries.

We store data in approximately its native form in the PostgreSQL database, using materialized views to extract components into normalized relations.
JSON data is stored as-is as \texttt{JSONB} objects, MARC records are stored in a relation with one entry per MARC field, etc.
The data tools package that accompanies this paper\footnote{\url{https://github.com/BoiseState/bookdata-tools}} provides links to download input data files, Python and Rust code to ingest raw data into the database, and SQL schema files to process and integrate the data sources in PostgreSQL.

\subsection{User Profiles and Book Ratings}

\begin{table}[tb]
    \centering
    \begin{tabular}{lrrrr}
\toprule
{} &     Users &     Items &       Pairs & Density \\
\midrule
AZ   & 8,026,324 & 2,269,633 &  22,464,284 & 0.0001\% \\
BX-E &    77,805 &   151,961 &     426,929 & 0.0036\% \\
BX-I &   105,283 &   279,983 &   1,128,965 & 0.0038\% \\
GR-E &   808,782 & 1,081,709 &  86,539,647 & 0.0099\% \\
GR-I &   870,011 & 1,097,581 & 188,962,927 & 0.0198\% \\
\bottomrule
\end{tabular}
    \caption{Interaction data summaries.}
    \label{tab:rating-summary}
\end{table}

We us three public sources of user-book interactions.  For each, we treat it both as an \emph{explicit feedback} data set by consulting rating values, and as an \emph{implicit feedback} data set by ignoring rating values and considering user-item interactions as positive signals.
In implicit-feedback settings, we consider all books a user has interacted with as positive implicit signals, even if they have a low rating: this corresponds to the signal that a bookseller can derive from sales data, as they do not know whether readers actually like the books they purchase once they have read them.

The \textbf{BookCrossing} (BX) data set \citep{Ziegler2005-zo} contains 1.1M book interactions from the BookCrossing reading community.
This data set contains both explicit ratings, on a 1--10 scale, and ``implicit'' actions of unspecified nature.
Since not all ratings have rating values, for explicit-feedback settings we exclude implicit actions, resulting in the ``BX-E'' data set; ``BX-I'' contains all BookCrossing interactions without rating values.

The \textbf{Amazon Books} (AZ) data set \citep{McAuley2015-rq} contains 22.5M reviews and ratings of books provided by customers on Amazon.com.
We use only the rating values, not the review text; since all recorded interactions have rating values, we use the interactions as-is and do not need to subset for explicit feedback.

The \textbf{GoodReads} (GR) data set \citep{monotonic} contains 189M interactions including ratings, reviews, and ``add to shelf'' actions from GoodReads, a reading-oriented social network and book discovery service.
As with BookCrossing, we extract a rating-only subset (``GR-E'') for explicit-feedback analysis, and use all user-book interactions (``GR-I'') for implicit feedback.

These data sets provide our historical user profiles (for RQ2) and the training data for our collaborative filtering algorithms.
All three are general reading data sets, consisting of user ratings for books across a wide range of genres and styles.
Table~\ref{tab:rating-summary} summarizes these data sets' basic statistics.
The ``Pairs'' column indicates the number of unique user-item pairs that appear in the data set.
We resolve multiple editions of the same work into a single item (see Section~\ref{sec:cluster}), so the item counts we report here may differ slightly from the item counts reported in other uses of these same rating data sets.

\subsection{Book Bibliographic Records}

We obtain book data, particularly author lists, by pooling records from Open Library\footnote{\url{https://openlibrary.org/developers/dumps}} and the Library of Congress (LOC) MARC Open-Access Records\footnote{\url{https://www.loc.gov/cds/products/marcDist.php}}.

We link these book records to rating data by ISBN.
Both OpenLibrary and LOC record ISBNs for book entries, and all book rating sources record ISBNs for the books users interact with (in the BookCrossing data, ISBN is the primary key for books; Amazon uses ISBNs as the identification numbers for books that have them).

Unfortunately, ISBN fields in the Library of Congress data are inconsistently formatted and used, including ISBNs in a range of formats as well as text other than ISBNs (many book entries store the cover price in the ISBN field).
We use a regular expression to look for sequences of 10 or 13 digits (allowing an X for the last digit in 10-digit sequences), optionally including spaces or hyphens, and treated those as ISBNs.  We do not validate check digits, preferring to maximize the ability to match ISBNs in the wild.

\subsection{ISBN Grouping}

In order to recommend at the level of creative works instead of individual editions, we group related ISBNs into a single ``item''.
This has the additional effects of decreasing the sparsity of the interaction data and increasing our data linking coverage.

To group ISBNs, we form a bipartite graph of ISBNS and record IDs.
Library of Congress bibliography records, OpenLibrary ``edition'' records, and GoodReads book records all constitute records for this purpose.
In addition, OpenLibrary and GoodReads each have a concept of a ``work''; when an edition or book is linked to a work, we use the work ID instead of the individual edition or book ID.

We then find the connected components on this graph, consider each component to be an ``item'', and assign it a single item identifier.

This process serves a similar purpose as ISBN linking services such as thingISBN \citep{thingisbn} and OCLC's xISBN service, but is completely reproducible using open data sources.

Rarely (less than 1\% of ratings) this causes a user to have multiple ratings for a book; we resolve multiple ratings in explicit-feedback settings by taking the median rating value.  Taking the most recent rating would also be a reasonable option, but BookCrossing does not include timestamps; since multiple ratings appear so infrequently, the precise strategy is unlikely to have significant impact on our results.

\subsection{Author Gender Data}

We obtain author information from the Virtual Internet Authority File (VIAF)\footnote{\url{http://viaf.org/viaf/data/}}, a directory of author information (\emph{Name Authority Records}) compiled from authority records from the Library of Congress and other libraries around the world.
Author gender identity (MARC Authority Field 375) is one of the available attributes for many records.

\subsubsection{Gender Identity Coding}

The MARC21 Authority Record data model \citep{Library_of_Congress1999-dv} employed by the VIAF is flexible in its ability to represent author gender identities, supporting an open vocabulary and begin/end dates for the validity of an identity.
The Program for Cooperative Cataloging provides a working group report on best practices for recording author gender identities, particularly for authors who are transgender or have a non-binary gender identity \citep{pcc-gender}.

Unfortunately, the VIAF does not use this flexibility --- all its gender identity records are ``male'', ``female'', or ``unknown''.
The result is that gender minorities are not represented, or are misgendered, in the available data.
We agree with \citet{Hoffmann2018-qt} that this is a significant problem.
The Library of Congress records better data, and as of August 2019 is in the process of preparing new exports of their linked data servies; we hope this will enable future research to better account for the complex nature of human gender identity and expression.

\subsubsection{Linking Author Data}

Because OpenLibrary, LOC, and VIAF do not share linking identifiers, we must link books to authority records by author name.
Each VIAF authority record can contain multiple name entries, recording different forms or localizations of the author's name.
OpenLibrary author records also carry multiple known forms of the author's name.
After normalizing names to improve matching (removing punctuation and ensuring both ``Last, First'' and ``First Last'' forms are available), we locate all VIAF records containing a name that matches one of the listed names for the first author of any OpenLibrary or LOC records in a book's ISBN group.
If all records that contain an assertion of the author's gender agree, we take that to be the author's gender; if there contradicting gender statements, we code the book's author gender as ``ambiguous''.

\begin{table}[tbp]
\centering
\begin{tabular}{lrrrrrrr}
\toprule
Data Set &  No Bk &  No Auth &  No VIAF &  Unknown &  Ambig. &  Male &  Female \\
\midrule
LOC     &     --- &          15.6\% &            6.0\% &    24.7\% &       0.9\% & 40.8\% &   12.0\% \\
AZ      &    46.2\% &           7.1\% &            6.7\% &    10.5\% &       0.7\% & 20.0\% &    8.8\% \\
BX-E    &    15.3\% &           4.3\% &            5.1\% &    11.6\% &       2.3\% & 36.5\% &   24.8\% \\
BX-I    &    16.7\% &           4.7\% &            5.7\% &    13.0\% &       2.1\% & 34.3\% &   23.6\% \\
GR-E    &     --- &          53.2\% &            2.7\% &     8.0\% &       0.8\% & 21.9\% &   13.3\% \\
GR-I    &     --- &          53.3\% &            2.7\% &     8.0\% &       0.8\% & 21.9\% &   13.2\% \\
\bottomrule
\end{tabular}
\caption{Summary of gender coverage (\% of books with each resolution result).}
\label{tbl:link-cover}
\end{table}

We selected this strategy to balance good coverage with confidence in classification.
Different authors with the same full name but different genders are unlikely to be a common occurrence.
Less than 2.5\% of rated books have `ambiguous' author genders.
Table~\ref{tbl:link-cover} shows relative frequency of link results for the books in our data sets; the columns correspond to the following failure points:

\begin{enumerate}
    \item \emph{No Bk} means the record book could not be linked to a book record of any kind. GoodReads has 100\% coverage since it comes with book records, but those records are not used for any data other than record identifiers.
    \item \emph{No Auth} means a book record was found, but had no authors listed.
    \item \emph{No VIAF} means authors were found, but none could be matched to VIAF.
    \item \emph{Unknown} means a VIAF record was found, but there were either no gender identity records or all records said ``unknown''.
    \item \emph{Ambiguous}, \emph{Male}, and \emph{Female} are the results of actual gender identity assertions.
\end{enumerate}

In the remainder of this paper, we group all no-data conditions together as ``unlinked''.

\subsubsection{Alternative Approaches to Author Gender}

Other work on understanding the behavior of computing systems with respect to gender and other demographic attributes that have been the basis of historic and/or ongoing discrimination uses various inference techniques to determine the demographics of data subjects.
This includes statistical detection based on names \citep{misloveUnderstandingDemographicsTwitter2011} and the use of facial recognition technology \citep{riedererPriceFairnessLocation2017}.

Such sources, however, are reductionistic \citep{hamidiGenderRecognitionGender2018} and often rely on and reinforce stereotypes regarding gender presentation.
Further, even to the extent that face-based gender recognition does work, it is biased in recognizing gender more accurately for lighter-skinned subjects \citep{gender-shades}.
We do not think inference techniques form a sound basis for understanding the social effects of computing systems.

Further, the Program for Cooperative Cataloging working group report specifically discourages inference of gender identity, even when the inference is performed by human, admonishing catalogers to ``not assume gender identity based on pictures or names'' \citep{pcc-gender}.
Catalogers following the recommendations learn an author's gender from explicit statements from official sources regarding their gender or the choice of pronouns or inflected nouns in official sources (such as the author's bio on the book cover).

\subsection{Data Set Statistics}

\begin{figure}[tb]
\includegraphics[width=\textwidth]{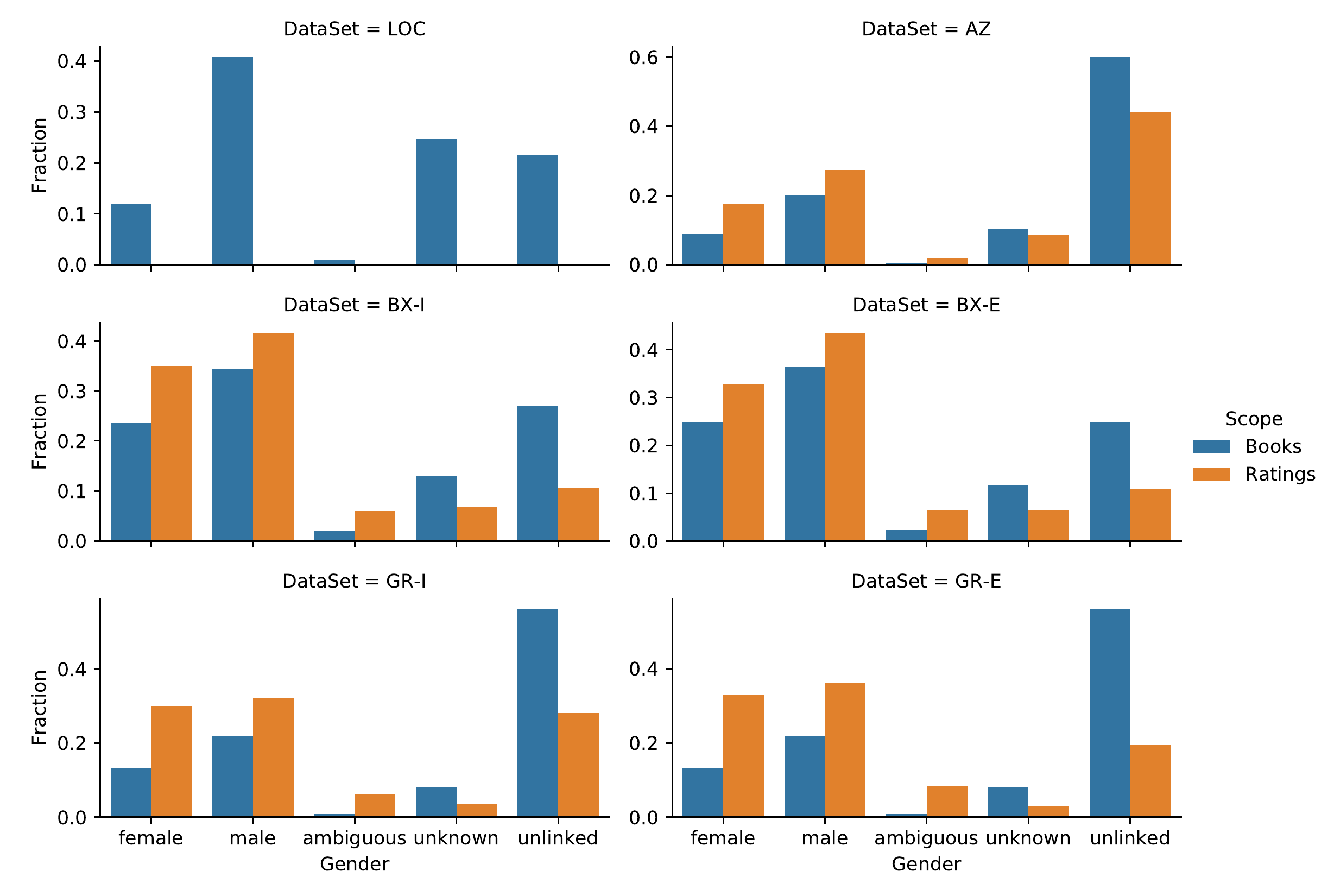}
\caption{Results of data linking and gender resolution. LOC is the set of books with Library of Congress records; other panes are the results of linking rating data.}
\label{fig:link-results}
\end{figure}

\begin{figure}
\includegraphics[width=0.5\textwidth]{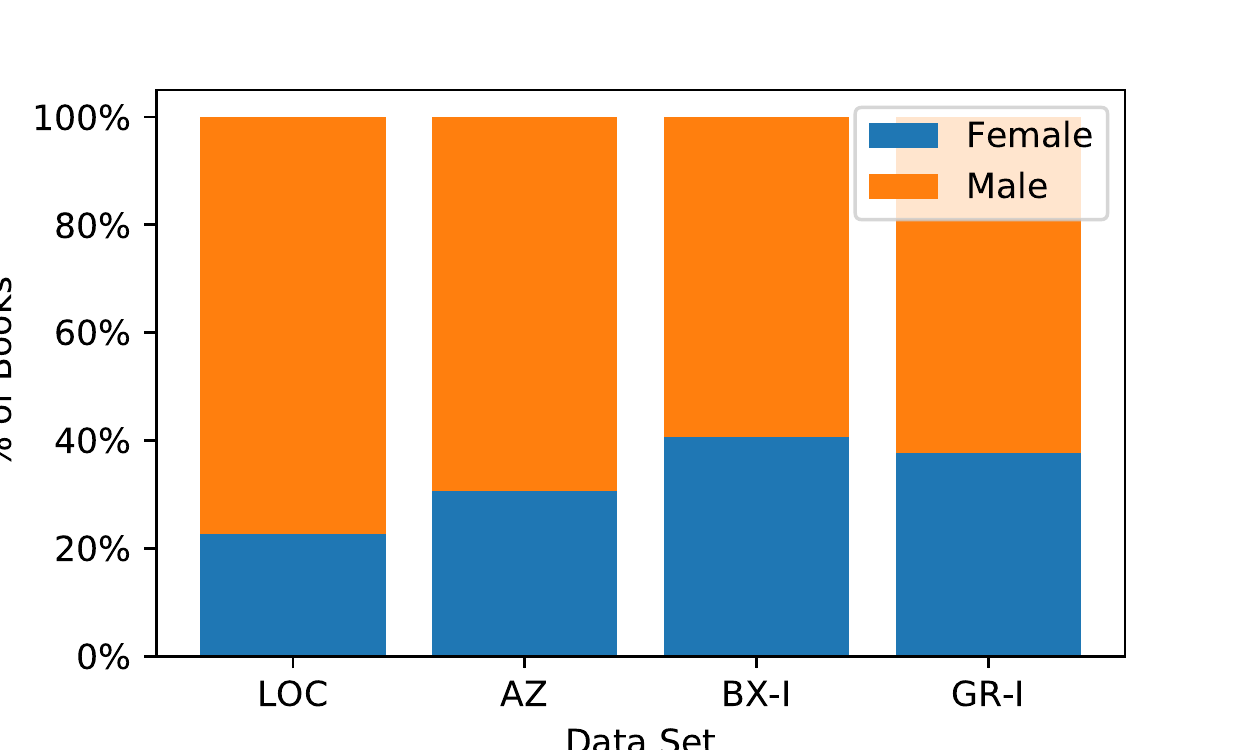}
\includegraphics[width=0.5\textwidth]{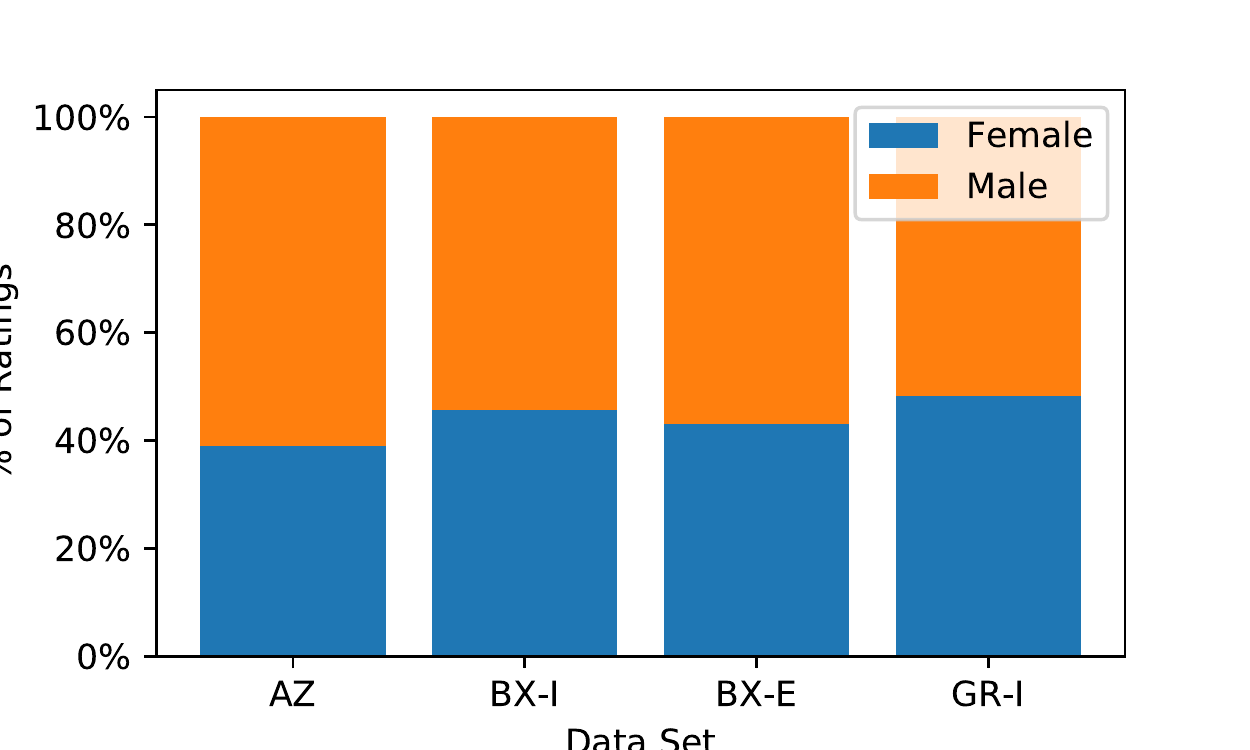}
\caption{Distribution of known-gender books in each data set.}
\label{fig:known-results}
\end{figure}

\begin{table}[tb]
    \centering
    \begin{tabular}{lrrrr}
\toprule
{} & \multicolumn{2}{c}{Books} & \multicolumn{2}{c}{Ratings} \\
Data & female &  male &  female &  male \\
\midrule
LOC     &  22.7\% & 77.3\% &   --- & --- \\
AZ      &  30.6\% & 69.4\% &   38.9\% & 61.1\% \\
BX-E    &  40.5\% & 59.5\% &   43.0\% & 57.0\% \\
BX-I    &  40.7\% & 59.3\% &   45.7\% & 54.3\% \\
GR-E    &  37.8\% & 62.2\% &   47.6\% & 52.4\% \\
GR-I    &  37.7\% & 62.3\% &   48.2\% & 51.8\% \\
\bottomrule
\end{tabular}
    \caption{Distribution of known-gender books and ratings.}
    \label{tab:known-dist}
\end{table}

Table \ref{tbl:link-cover} and Fig. \ref{fig:link-results} summarize the results of integrating these data sets. 
While the data is sparse, it has sufficient coverage for us to perform a meaningful analysis.
We also report coverage of the Library of Congress data itself, as a rough approximation of books published irrespective of whether they are rated.
Unfortunately, we do not know what biases lie in the coverage rates: are unlinked or unknown books more likely to be written by authors of one gender or another?

\subsection{RQ1: Baseline Corpus Distribution}

This analysis, and the distribution of genders show in in Fig. \ref{fig:known-results} and Table \ref{tab:known-dist}, provide our answer to RQ1.
Of Library of Congress books with known author genders, 22.7\% are written by women.
Rating data sets have higher representation of women: 30.6\% of books rated on Amazon are written by women, and 40.7\% of BookCrossing books.
Representation is higher yet when looking at ratings themselves: while 37.7\% of known-gender books on GoodReads are written by women, 48.2\% of shelf adds of known-gender books are for books by women.

If women are underrepresented in book publishing, they are less underrepresented in book rating data.
The GoodReads community achieves close to gender parity in terms of books rated or added to shelves.
\section{Experiment and Analysis Methods}

Starting with the integrated book data, our main experiment has several steps:

\begin{enumerate}
\item Sample 1000 users, each of whom has rated at least 5 books with known author gender, for analysis. 
\item Quantify gender distribution in sample user profiles (RQ2).
\item Produce 50 recommendations for each sample users, using the entire data set data set for training.
\item Compute recommendation list gender distribution (RQ3) and compare with user profile distribution (RQ4).
\end{enumerate}

\subsection{Sampling}

We sample 1000 users to keep the final data set tractable.
As will be seen, our statistical analysis methods are computationally intensive, scaling in the number of users.
Sampling users to use in assessing user profile makeup and gender propagation enables this analysis to be done in reasonable time; 1000 users is enough to ensure some statistical validity.

We require each user to have at least 5 books with known author gender so that their profile has enough books to estimate user gender balance, and so that the recommender has history with which to make recommendations.

\subsection{Recommending Books}
\label{sec:algorithms}

We used the LensKit toolkit~\citep{lkpy} to produce 50 recommendations for each of our 1000 sample users using the following algorithms:

\begin{itemize}
\item UU, a user-based collaborative filter \citep{Herlocker1999-pw}.  In implicit-feedback mode, it sums user similarities instead of computing a weighted average.
\item II, an item-based collaborative filter \citep{Deshpande2004-ar}.  As with UU, in implicit feedback mode, this algorithm sums user similarities instead of computing a weighted average.
\item ALS, a matrix factorization model trained with alternating least squares \citep{pilaszyFastAlsbasedMatrix2010}; we use both implicit and explicit feedback versions.
\item BPR, a learning-to-rank algorithm that optimizes pairwise ranking \citep{rendleBPRBayesianPersonalized2009}; we use the BPR-MF version.
\end{itemize}

\subsubsection{Tuning and Performance}
\label{sec:perf}

\begin{figure}[tb]
    \centering
    \includegraphics[width=\textwidth]{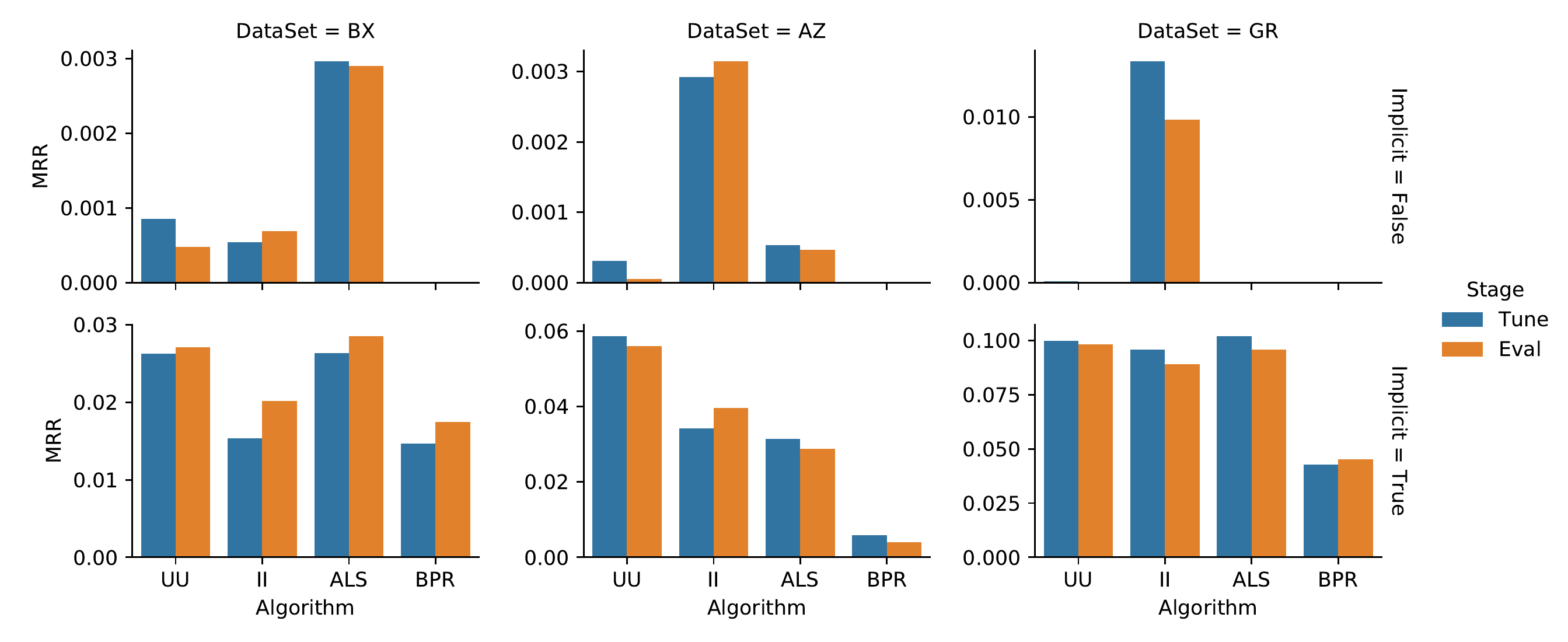}
    \caption{Top-$N$ recommendation accuracy.}
    \label{fig:acc-base}
\end{figure}

While recommendation accuracy is not the focus of our experiment, we report it for context; it also provides a baseline for our exploration of distribution-constraining rerankers in Section~\ref{sec:force}.

We sampled 5000 users with at least 5 ratings for evaluation.  For each user, we held out one rating as the test rating, generated a 100-item recommendation list, and measured the Mean Reciprocal Rank (MRR).

We tuned each model's hyperparameters with scikit-optimize, optimizing MRR on a separate tuning set that was selected identically to the evaluation set.\footnote{To reduce the number of zeros, we tuned GoodReads using 1000-item lists instead of 100.}
We used scikit-optimize's Gaussian process minimization routine and stopped when the 5 best settings showed no more than 1\% improvement in MRR.

We exclude ALS on GR-E because it did not perform well after repeated tuning attempts. Implicit ALS worked well on GoodReads.

\begin{table}[tb]
    \centering
\begin{tabular}{lrrrr}
\toprule
Data Set  &      ALS &       BPR &        II &        UU \\
\midrule
AZ (E) &  0.000474 &      --- &  0.002507 &  0.000333 \\
AZ (I) &  0.029904 &  0.002615 &  0.032598 &  0.050321 \\
BX-E &  0.002636 &       --- &  0.000547 &  0.000760 \\
BX-I &  0.029206 &  0.015749 &  0.019148 &  0.025843 \\
GR-E &       --- &       --- &  0.010657 &  0.000066 \\
GR-I &  0.105331 &  0.048478 &  0.096841 &  0.105628 \\
\bottomrule
\end{tabular}
    \caption{Evaluation MRR.}
    \label{tab:eval-mrr}
\end{table}

Fig. \ref{fig:acc-base} shows the MRR both on the evaluation set and on the tuning set with the best hyperparameters, and Table \ref{tab:eval-mrr} lists the final evaluation MRR values.
Nearest-neighbor recommenders performed quite well on implicit-feedback data; we suspect this is partially due to popularity bias \citep{belloginPrecisionorientedEvaluationRecommender2011}, as similarity-sum implicit-feedback k-NN will favor popular items.

\subsection{Statistical Analysis}

Our statistical goal is to estimate the gender balance of user profiles, recommendation lists, and the propagation factor between them.
There are several challenges that complicate doing this with commonly-used statistical techniques:

\begin{itemize}
    \item Variance in user profile sizes makes it difficult to directly compare gender proportions between users (2 out of 5 and 20 out of 50 reflect very different levels of confidence).
    \item With many data sets and algorithm, we quickly run into large (and non-obvious) multiple comparison problems.
    \item We are interested in assessing distributions of bias, not just point estimates.
\end{itemize}

\begin{figure}
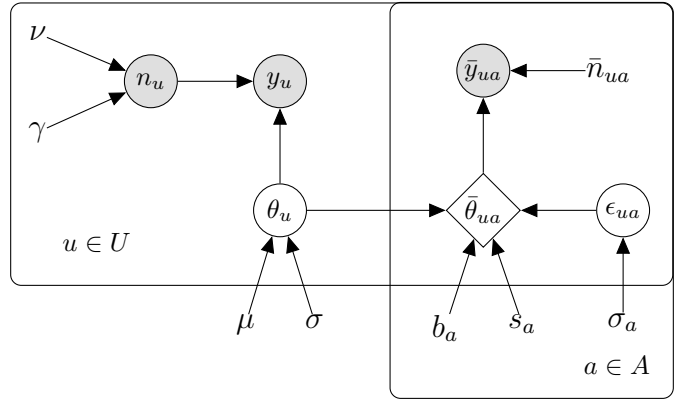

\centering
\tikz{ %
\node[latent] (theta) {$\theta_u$} ; %
\node[obs, above=of theta] (y) {$y_u$} ; %
\node[obs, left=of y] (n) {$n_u$} ; %
\node[const, above left=of n, yshift=-1em, xshift=-1em] (nu) {$\nu$} ;
\node[const, below left=of n, yshift=1em, xshift=-1em] (gamma) {$\gamma$};
\node[const, below left=of theta, yshift=-1em, xshift=1.5em] (mu) {$\mu$};
\node[const, below right=of theta, yshift=-1em, xshift=-1.5em] (sigma) {$\sigma$};

\edge {theta} {y} ; %
\edge {n} {y} ; %
\edge {nu} {n} ;
\edge {gamma} {n} ;
\edge {mu} {theta};
\edge {sigma} {theta};

\node[det, right=of theta, xshift=2em] (thetaL) {$\bar\theta_{ua}$};
\node[latent, right=of thetaL] (epL) {$\epsilon_{ua}$};
\node[obs, above=of thetaL] (yL) {$\bar y_{ua}$};
\node[const, right=of yL] (nL) {$\bar n_{ua}$};
\node[const, below left=of thetaL, yshift=-1em, xshift=1.5em] (ba) {$b_a$};
\node[const, below right=of thetaL, yshift=-1em, xshift=-1.5em] (sa) {$s_a$};
\node[const, below=of epL] (siga) {$\sigma_a$};

\edge {theta} {thetaL};
\edge {epL} {thetaL};
\edge {thetaL} {yL};
\edge {nL} {yL};
\edge {ba} {thetaL};
\edge {sa} {thetaL};
\edge {siga} {epL};

\node[caption, left=of theta, xshift=-3.5em, yshift=-1em] (ulab) {$u \in U$};
\node[draw, rectangle, rounded corners, 
	fit=(ulab) (theta) (y) (n) (thetaL) (epL) (yL) (nL), inner sep=1em, xshift=-0.12cm, yshift=0.12cm] {};

\plate[inner sep=1em, xshift=-0.12cm, yshift=0.12cm] {algoPlate} {(ba) (sa) (siga) (thetaL) (epL) (yL) (nL)} {$a \in A$};
}
\caption{Plate diagram for statistical model.}
\label{fig:model-plate}
\end{figure}

\begin{table}[tb]
\caption{Summary of key model parameters and variables.}
\label{tbl:variables}
\centering
{\small
\begin{tabular}{rp{2.7in}}
Variable & Description \\
\toprule
$n_u$ & Number of known-gender books rated by user $u$ \\
$y_u$ & Number of female-authored books rated by $u$ \\
$\theta_u$ & Probability of a known-author book rated by $u$ being by a female author (smoothed author-gender balance) \\
\midrule
$\mu$ & Expected user gender balance, in log-odds ($\mathrm{E}[\mathrm{logit}(\theta_u)]$) \\
$\sigma^2$ & Variance of user gender balance \\
\midrule
$\bar n_{ua}$ & Number of known-gender books algorithm $a$ recommended to user $u$ \\
$\bar y_{ua}$ & Number of female-authored books $a$ recommended to $u$ \\
$\bar\theta_{ua}$ & Gender balance of algorithm $a$'s recommendations for $u$ \\
\midrule
$s_a$ & Regression slope of algorithm $a$ (its responsiveness to user profile tendency) \\
$b_a$ & Intercept of algorithm $a$ \\
$\sigma_a^2$ & Residual variance of algorithm $a$ (its variability unexplained by user tendencies) \\
\bottomrule
\end{tabular}}
\end{table}

To address these difficulties, we model user rating behaviors using a hierarchical Bayesian model~\citep{Gelman2014-iq} for the observed number of books by female authors out of the set of books with known authors.
This model allows us to integrate information across users to estimate a user's tendency even when they have not rated very many books, and integrated Bayesian models enable us to robustly infer a number of parameters in a manner that clearly quantifies uncertainty and avoids many multiple-comparison problems \citep{Gelman2000-bj}.

We extend this to model recommendation list distributions as a linear function of user profile distributions plus random variance.

Figure~\ref{fig:model-plate} shows the plate diagram for our model, and Table~\ref{tbl:variables} summarizes the key parameters.

\subsubsection{User Profiles}
\label{sec:model-profile}
For each user, we observe $n_u$, the number of books they have rated with known author gender, and $y_u$, the number of female-authored books they have rated.
From these observations, we estimate each user's author-gender tendency $\theta_u$ using a logit-normal model to address RQ2.
The beta distribution is commonly used for modeling such tendencies, but the logit-normal has two key advantages: it is more parsimonious when extended with a regression, as we can compute regression coefficients in log-odds space, and it is substantially more computationally efficient to sample.
In early versions of this experiment we also found that it fit our data slightly better.

We also model $n_u$ as a random variable with a negative binomial distribution so that we can produce realistic predicted observations for unseen users to test model fit (by comparing the distribution of actual observed user rating counts to those of synthetic users drawn from the posterior distribution).

We use the following joint probability as our likelihood model:

\begin{align*}
y_u & \sim \mathrm{Binomial}(n_u, \theta_u) \\
\mathrm{logit}(\theta_u) & \sim \mathrm{Normal}(\mu, \sigma) \\
n_u & \sim \mathrm{NegBinomial}(\nu,\gamma)
\end{align*}

$\mathrm{logit}(\theta_u)$ is the log odds of a known-gender book rated by user $u$ being written by a female author, and $\mu$ and $\sigma$ are the mean and standard deviation of this user author-gender tendency.  
Negative values indicate a tendency towards male authors, and positive values a tendency towards female authors. 
$\theta_j$ is the corresponding probability or proportion in the range $[0,1]$.
When sampling from the fitted model, we produce a predicted $\theta'$, $n'$, $y'$, and observed ratio $y'/n'$ for each sample in order to estimate the distribution of unseen user profiles.

We put vague priors on all parameters: $\sigma, \nu, \gamma \sim \mathrm{Exponential}(0.01)$, as they are positive, and $\mu \sim \mathrm{Normal}(0, 100)$.
These priors provide diffuse density across a wide range of plausible and extreme values.

\subsubsection{Recommendation Lists}
\label{sec:model-reclists}
For RQ3 and RQ4, we model recommendation list gender distributions by extending our Bayesian model to predict recommendation distributions with a linear regression based on each user's smoothed proportion and per-algorithm slope, intercept, and variance.
The regression is in log-odds (logit) space, and results in the following formula for estimating $\bar\theta_{ua}$:

\begin{align*}
\mathrm{logit}(\bar\theta_{ua}) & = b_a + s_a \mathrm{logit}(\theta_u) + \epsilon_{ua} \\
\epsilon_{ua} & \sim \mathrm{Normal}(0, \sigma_a) \\
\end{align*}

While it would seem logical to use a binomial distribution to model observed recommendation list author counts, in practice it does not work as well as we would hope.
Recommendations are not independent between users, and consistency in recommendations cause a binomial model to fit poorly (observed proportions are severely underdispersed).
We therefore  omit the binomial distribution and directly compute $\bar\theta_{ua} = (\bar y_{ua} + 1) / (\bar n_{ua} + 2)$.
The regression residual $\epsilon_{ua}$ captures variance in the relationship between users' and algorithms' recommendation proportions, and giving it per-algorithm variance allows for some algorithms being more consistent in their output than others.

The final result of our full model is that $s_a$ captures how much an algorithm's output gender distribution varies \emph{with} the input profile distribution, and $\sigma_a^2$ its variance \emph{independent of} the input distribution.
$b_a$ expresses the algorithm's typical gender balance when the user's profile is evenly balanced (since the log of even odds is zero).

\subsubsection{Implementation}
We fit and sample models with STAN \citep{Carpenter2017-gq}, drawing 10,000 samples per model (4 NUTS chains each performing 2500 warmup and 2500 sampling iterations).
We report results with the posterior predictive distributions of the parameters of interest, as estimated by the sampling process.
\section{Profile and Propagation Results}
\label{sec:results}

In this section we present the results of our statistical analysis of user profiles and recommendations.
We begin with characterizing the profiles of our sample users, and then proceed to analyze the resulting recommendations.

\subsection{User Profile Characteristics}
\label{sec:profiles}
\label{sec:RQ2}

\begin{figure}[tb]
\includegraphics[width=\columnwidth]{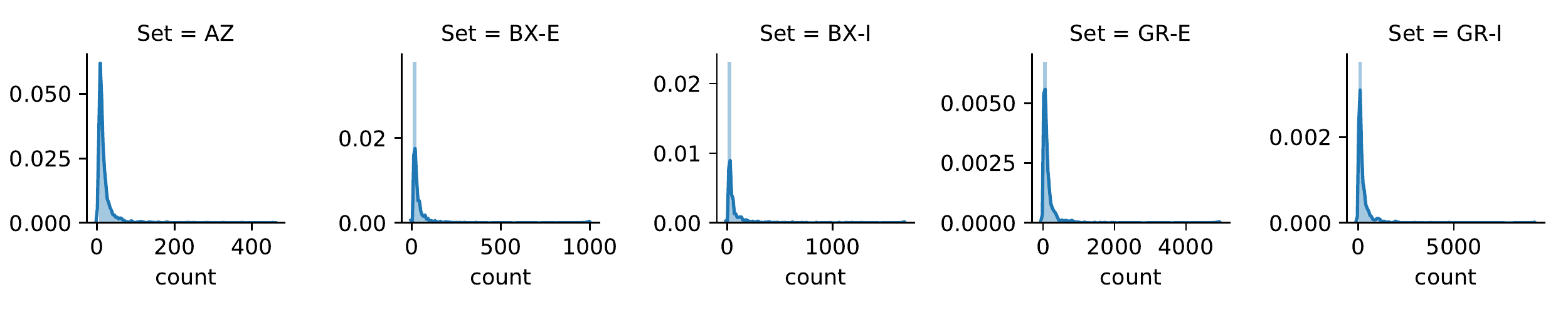}
\includegraphics[width=\columnwidth]{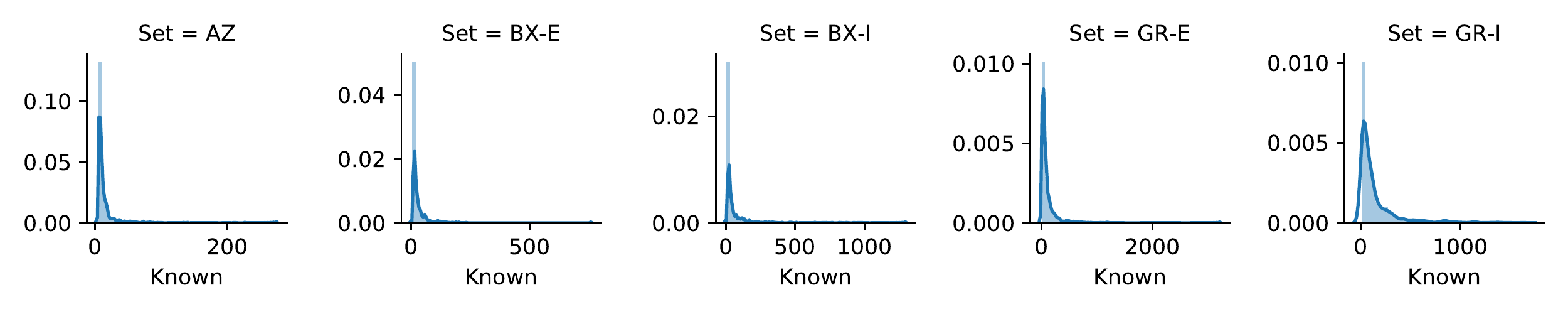}
\caption{Distribution of user profile sizes, of all books (top) and known-gender books (bottom).}
\label{fig:profile-size}
\end{figure}

\begin{table}[tb]
\caption{Summary statistics for user profile gender distributions (log odds of $P(\mathrm{female}|\mathrm{known})$).}
\label{tbl:profile-summary}
\centering{\small
\begin{tabular}{lccccc}
& AZ & BX-E & BX-I & GR-E & GR-I \\
\toprule
Mean Obs. Proportion & 0.394 & 0.412 & 0.410 & 0.453 & 0.448 \\
\hspace{1em}{\emph{Std. Dev.}} & 0.319 & 0.265 & 0.253 & 0.279 & 0.265 \\
\midrule
$\mu$ (est. mean log odds) & -0.63 & -0.44 & -0.43 & -0.26 & -0.28 \\
\hspace{1em}{\emph{95\% Interval for $\mu$}} & (-0.76, -0.51) & (-0.52, -0.35) & (-0.51, -0.35) & (-0.36, -0.15) & (-0.37, -0.19) \\
$\sigma$ (est. sd log odds) & 1.77 & 1.15 & 1.06 & 1.52 & 1.40 \\
\midrule
Posterior Mean $\theta$ & 0.40 & 0.41 & 0.42 & 0.46 & 0.45 \\
\hspace{1em}{\emph{Std. Dev.}} & 0.29 & 0.23 & 0.22 & 0.27 & 0.26\\
\bottomrule
\end{tabular}}
\end{table}

\begin{figure}[tbp]
\includegraphics[width=\columnwidth]{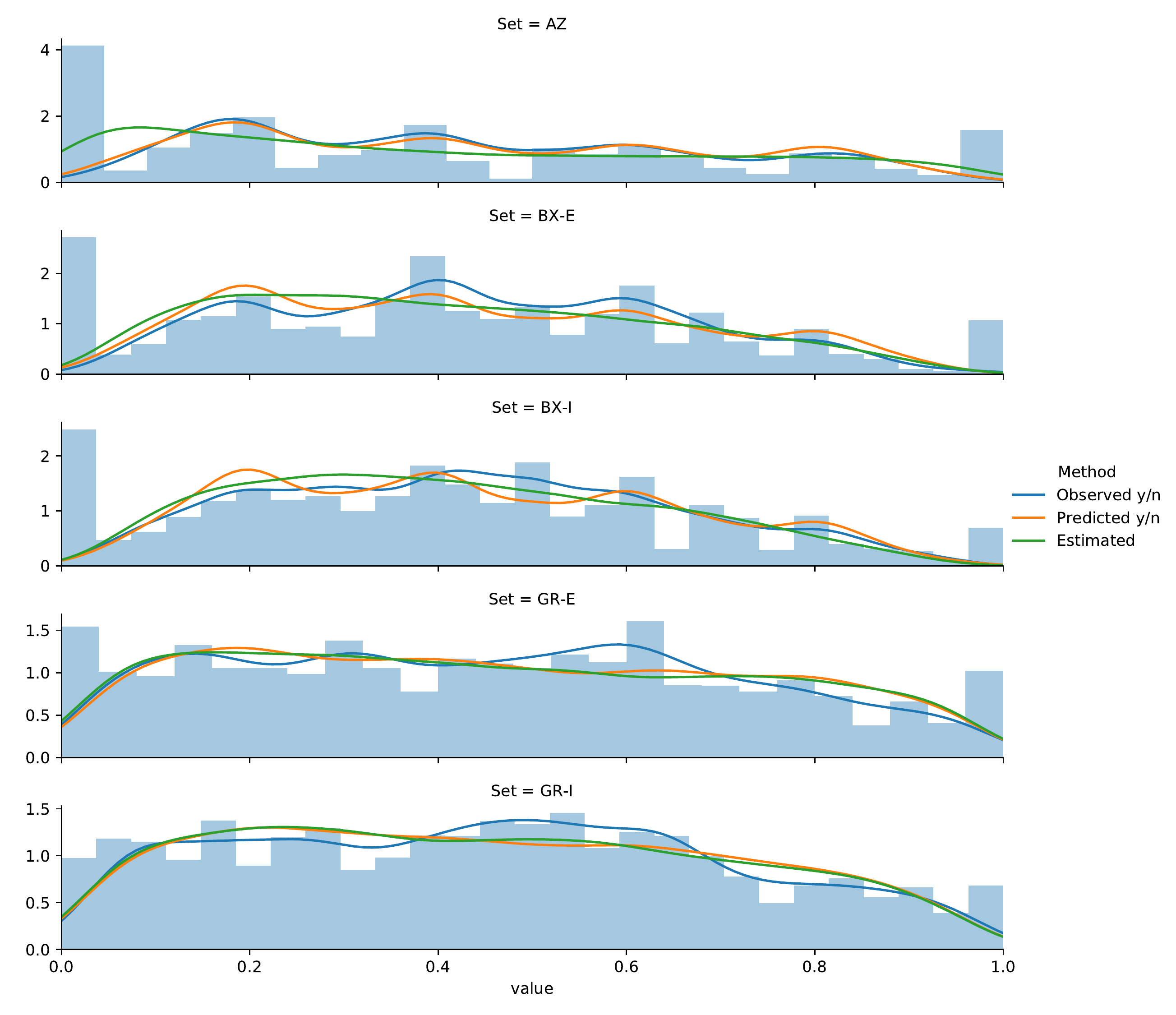}
\caption{Distribution of user author-gender tendencies.  Histogram shows observed proportions; lines show kernel densities of estimated tendencies ($\theta'$) along with observed and predicted proportions.}
\label{fig:profile-density}
\end{figure}

Under RQ2, we want to understand the distribution of users' author-gender tendencies, as represented by the proportion of known-gender books in each author's profile that are written by female authors.
Figure~\ref{fig:profile-size} shows the distribution of user profile sizes, and Figure~\ref{fig:profile-density} shows the distribution of observed author gender proportions.
Table~\ref{tbl:profile-summary} presents user profile summary statistics.

The Bayesian model from Section~\ref{sec:model-profile} provides more rigorous, smoothed estimates of this distribution.
Table~\ref{tbl:profile-summary} describes the numerical results of this inference.
The key parameters are $\mu$, the average user's author-gender tendency in log-odds; $\sigma$, the standard deviation of user author-gender tendencies; and sampled $\theta$ values, the distribution of which describes the distribution of user author-gender tendencies expressed as expected proportions.

Figure~\ref{fig:profile-density} shows the densities of the author-gender tendency distribution, along with the densities of projected and actual observed proportions.  The ripples in predicted and observed proportions are due to the commonality of 5-item user profiles, for which there are only 6 possible proportions; estimated tendency ($\theta$) smooths them out. This smoothing, along with avoiding estimated extreme biases based on limited data, are why we find it useful to estimate tendency instead of directly computing statistics on observed proportions.  
To support direct comparison of the densities of observations and predictions, we resampled observed proportions with replacement to yield 10,000 observations.

We observe a population tendency to rate male authors more frequently than female authors in all data sets ($\mu < 0$), but to rate female authors more frequently than they would be rated were users drawing books uniformly at random from the available set.
The average user author-gender tendency is slightly closer to an even balance than the set of rated books.
We also found a large diversity amongst users about their estimated tendencies (s.d. of predicted $\theta$ exceeds 0.2; inferred $\sigma > 1$; both even-odds and book population proportions are within one s.d. of estimated means).
This means that some users are estimated to strongly favor female authored books, even if these users are outnumbered by those that primarily read male-authored books.
The Amazon data set has the strongest tendency ($\mu=-0.63$, $\mathrm{mean}(\theta)=0.40$, $\mathrm{sd}(\theta)=0.29$), with a particular spike in highly-male profiles.

\subsection{Recommendation List Distributions}

\begin{table*}[tb]
\caption{Recommendation coverage and diversity statistics (implicit).}
\label{tbl:cov-imp}
\centering{\small
\begin{tabular}{lrrrrrrrrr}
\toprule
 & \multicolumn{3}{c}{AZ} & \multicolumn{3}{c}{BX} & \multicolumn{3}{c}{GR} \\
 &     Recs & Dist.  & \% Dist
 &     Recs & Dist.  & \% Dist
 &     Recs & Dist.  & \% Dist \\
\midrule
ALS       
&  50000 &  10920 &      21.8
&  50000 &   5926 &  11.9
&  50000 &   8783 &  17.6 \\
BPR       
&  50000 &   8764 &  17.5 
&  50000 &  19595 &  39.2
&  50000 &  35683 &  71.4 \\
II
&  49952 &  36582 &  73.2
&  49900 &  20417 &  40.9
&  50000 &   8569 &  17.1 \\
UU
&  49952 &  16287 &  32.6 
&  49660 &   5921 &  11.9
&  49908 &   6420 &  12.8 \\
\bottomrule
\end{tabular}}
\end{table*}

\begin{table*}[tb]
\caption{Recommendation coverage and diversity statistics (explicit).}
\label{tbl:cov-exp}
\centering{\small
\begin{tabular}{lrrrrrrrrr}
\toprule
 & \multicolumn{3}{c}{AZ} & \multicolumn{3}{c}{BX} & \multicolumn{3}{c}{GR} \\
 &     Recs & Dist.  & \% Dist
 &     Recs & Dist.  & \% Dist
 &     Recs & Dist.  & \% Dist \\
\midrule
ALS       
&  50000 &  10920 &      32.3
&  50000 &   69 &  0.1
&  --- &   --- &  --- \\
II
&  48326 &  35236 &  72.9
&  47941 &  10238 &  20.6
&  49534 &   19452 &  59.5 \\
UU
&  38417 &  31834 &  82.9 
&  44353 &   19570 &  44.1
&  31834 &   25547 &  52.6 \\
\bottomrule
\end{tabular}}
\end{table*}

\begin{figure*}
\includegraphics[width=\textwidth]{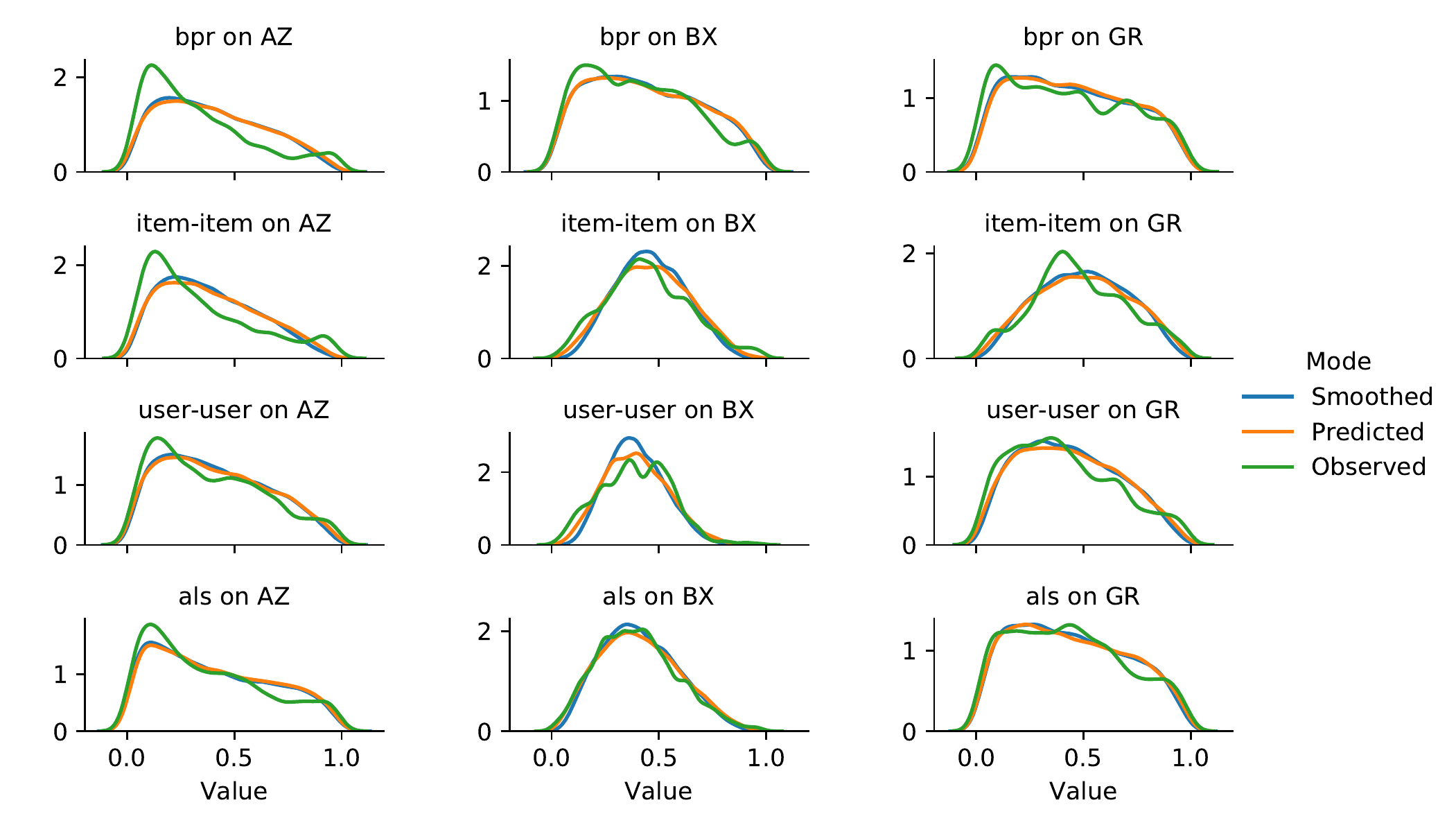}
\caption{Posterior densities of recommender biases from integrated regression model (implicit feedback).}
\label{fig:rec-densities-imp}
\end{figure*}

\begin{figure*}
\includegraphics[width=\textwidth]{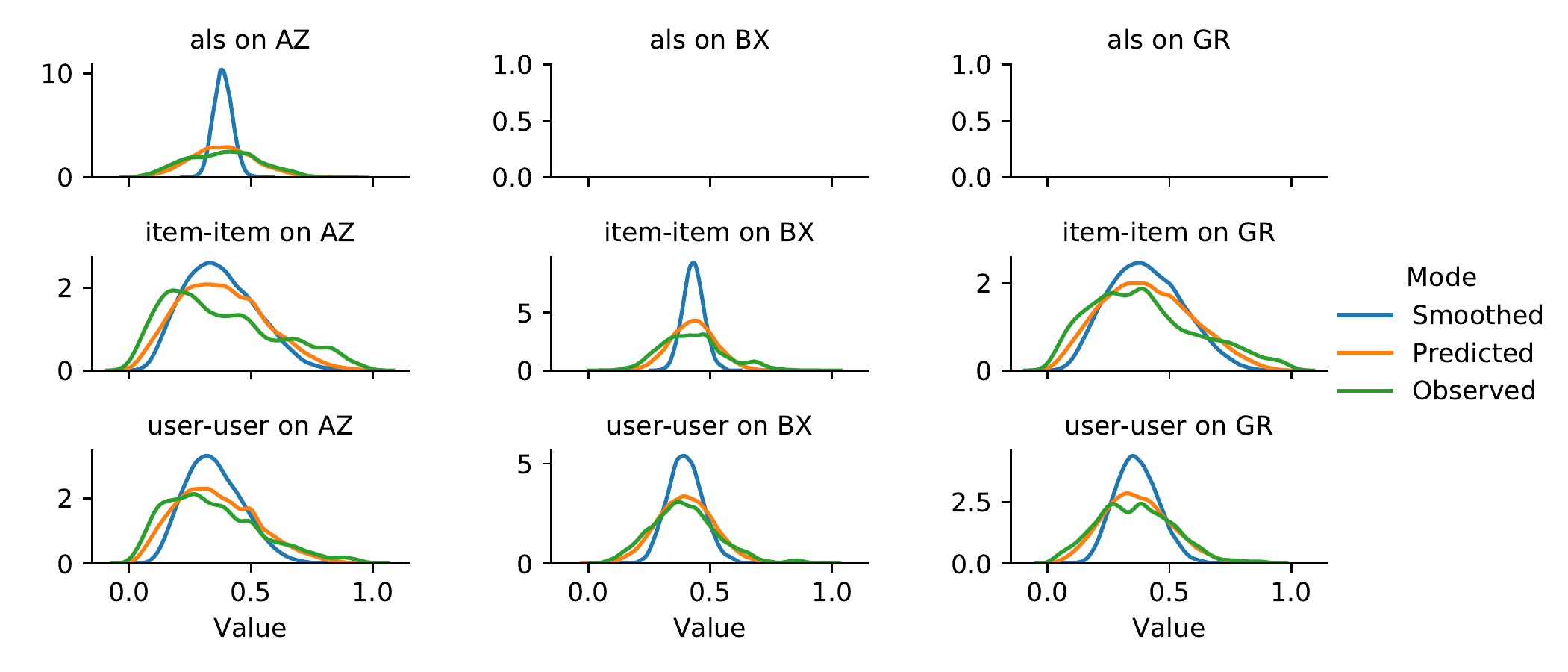}
\caption{Posterior densities of recommender biases from integrated regression model (explicit feedback).}
\label{fig:rec-densities-exp}
\end{figure*}

\begin{table}[bt]
\caption{Mean / SD of rec. list female author proportions.}
\label{tbl:rec-stats}
\centering{\small
\begin{tabular}{llccc}
& & AZ & BX & GR \\
\toprule
\multicolumn{2}{l}{Popular} & 0.486 & 0.390 & 0.424 \\
\midrule
\multirow{4}{*}{Implicit} & ALS
& 0.398 / 0.302 & 0.406 / 0.190 & 0.437 / 0.274
\\
& BPR
& 0.383 / 0.298 & 0.425 / 0.269 & 0.446 / 0.316 \\
& II
& 0.365 / 0.305 & 0.454 / 0.204 & 0.491 / 0.238
\\
& UU
& 0.401 / 0.271 & 0.393 / 0.169 & 0.422 / 0.260
\\
\midrule
\multirow{3}{*}{Explicit}
& ALS
& 0.380 / 0.155 & 0.286 / 0.010 & ---
\\
& II
& 0.378 / 0.243 & 0.430 / 0.130 & 0.400 / 0.239
\\
& UU
& 0.334 / 0.221 & 0.394 / 0.154 & 0.364 / 0.166
\\
\bottomrule
\end{tabular}}
\end{table}

\begin{figure*}
\includegraphics[width=0.8\textwidth]{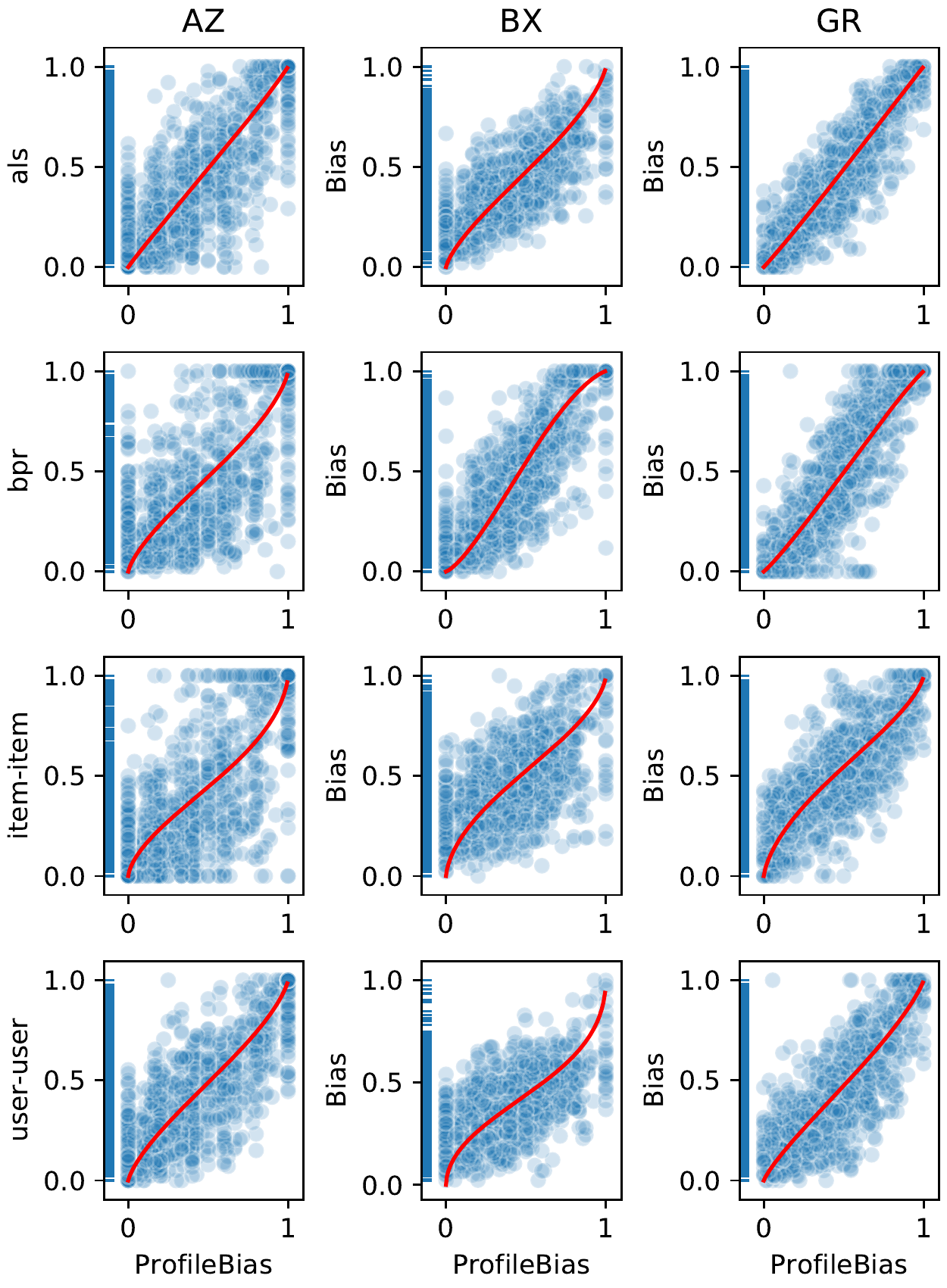}
\caption{Scatter plots and regression curves for implicit feedback recommender response to individual users.}
\label{fig:rec-regressions-imp}
\end{figure*}

\begin{figure*}
\includegraphics[width=0.8\textwidth]{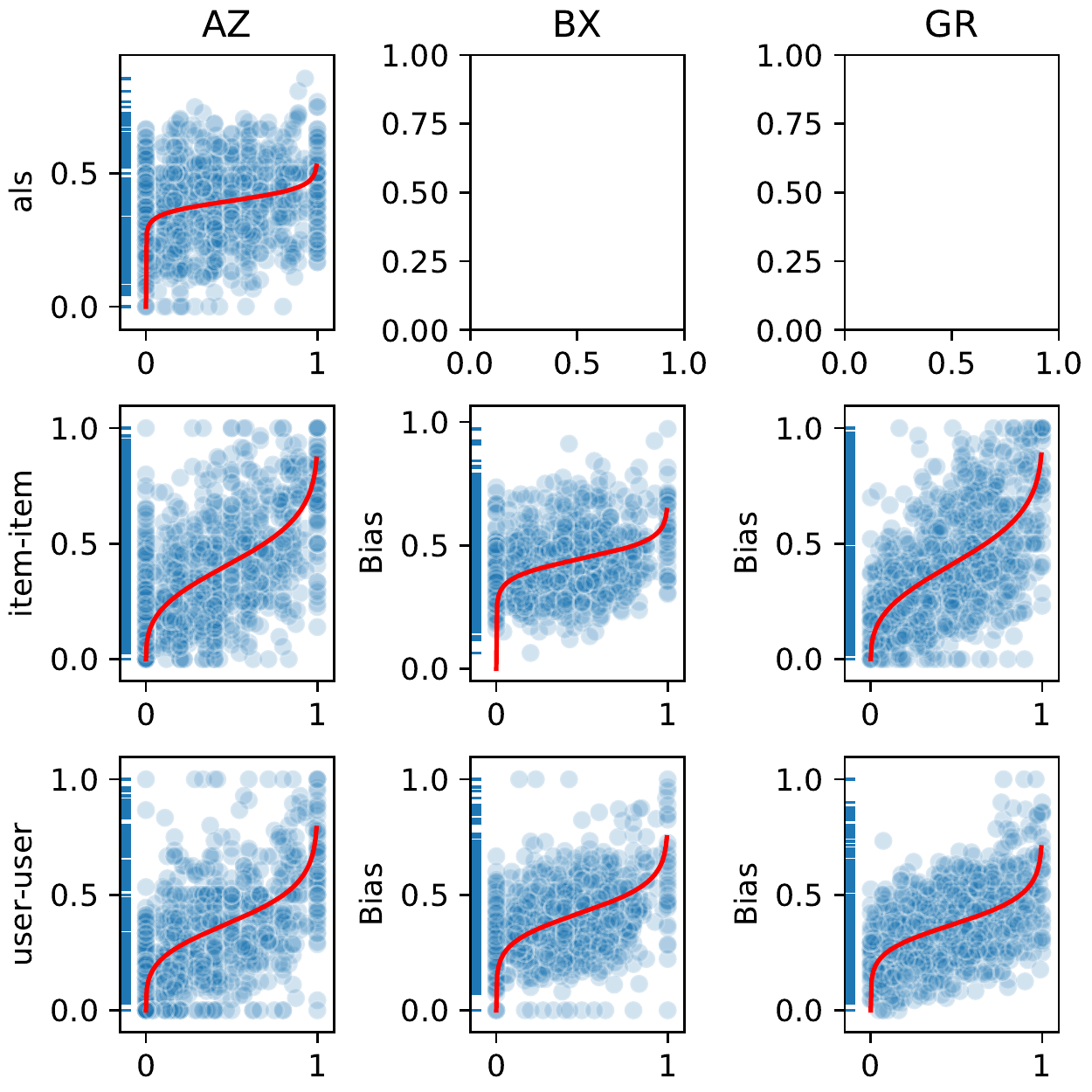}
\caption{Scatter plots and regression curves for explicit feedback recommender response to individual users.}
\label{fig:rec-regressions-exp}
\end{figure*}

\begin{figure*}
    \centering
\includegraphics[width=0.8\textwidth]{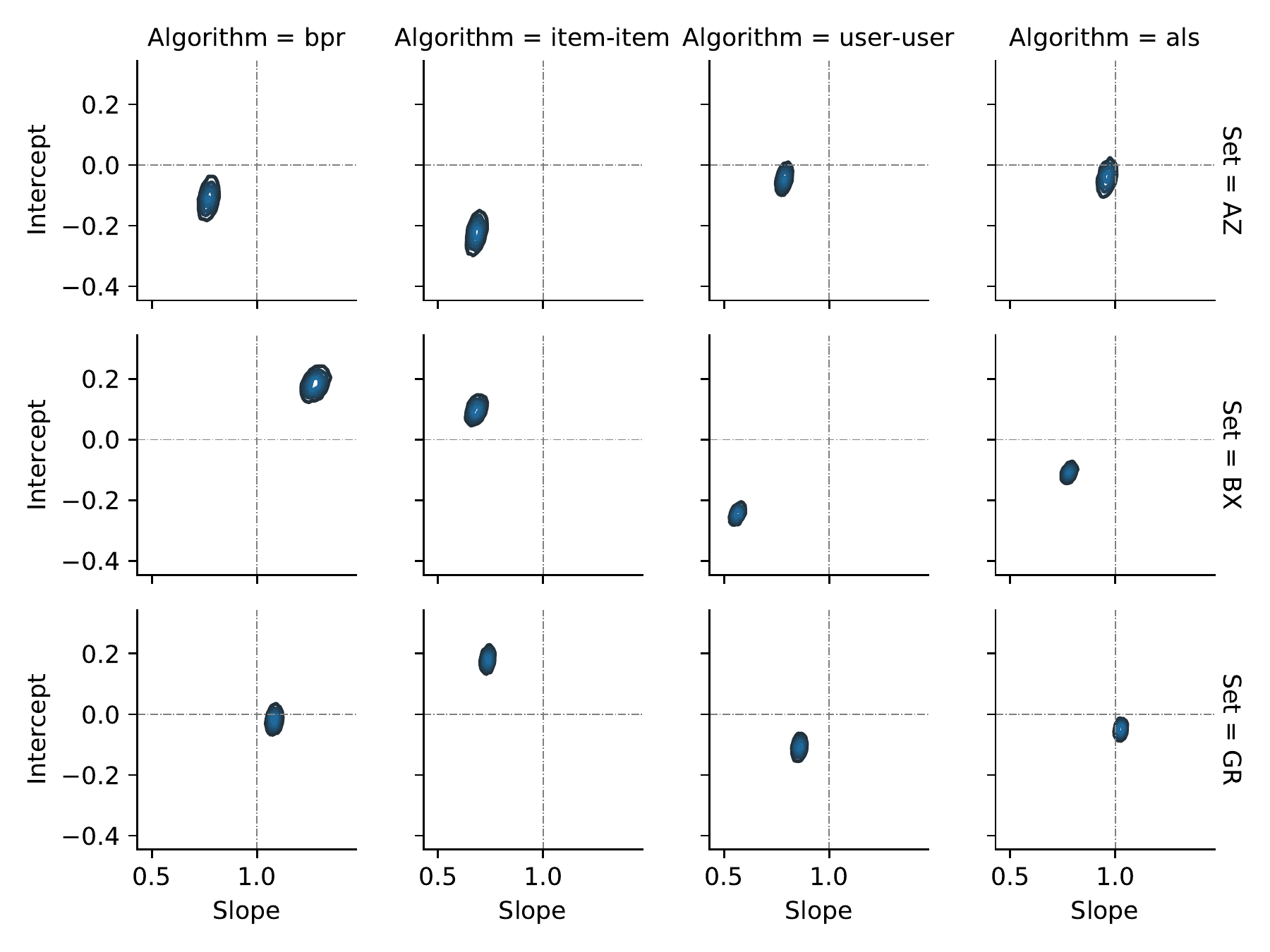}
\caption{Slope and intercept contour plots (implicit feedback).}
\label{fig:reg-contour-imp}
\end{figure*}

\begin{figure*}
    \centering
\includegraphics[width=0.8\textwidth]{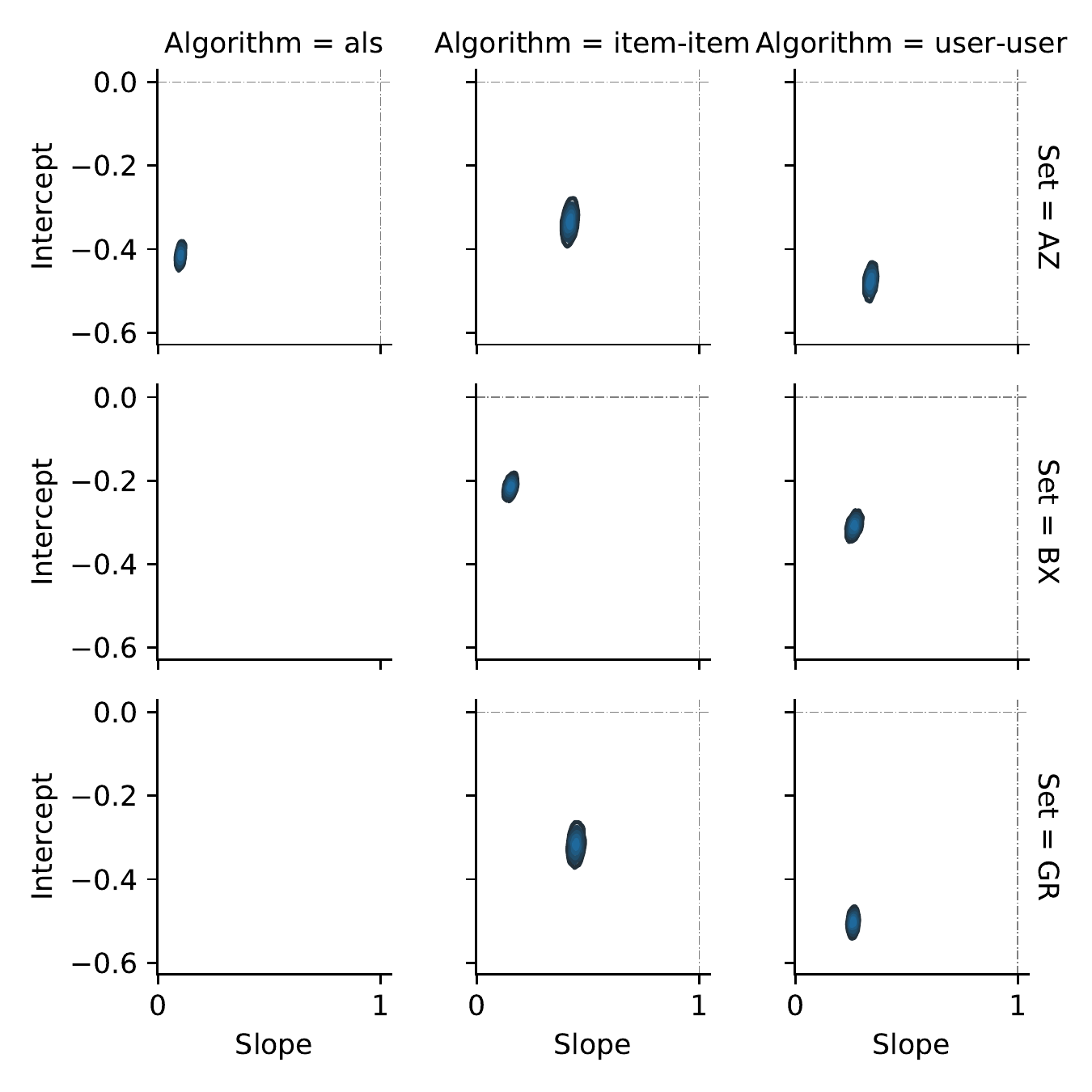}
\caption{Slope and intercept contour plots (explicit feedback).}
\label{fig:reg-contour-exp}
\end{figure*}

Our first step in understanding how collaborative filtering algorithms respond to this data bias is to examine the distribution of recommender list tendencies (RQ3).  As described in \ref{sec:algorithms}, we produced 50 recommendations from each algorithm.  Tables \ref{tbl:cov-imp} and \ref{tbl:cov-exp} show the basic coverage statistics of these algorithms. 
Users for which an algorithm could not produce recommendations are rare.
We also computed the extent to which algorithms recommend different items to different users; ``\% Dist.'' is the percentage of all recommendations that were distinct items.
Algorithms that repeatedly recommend the same items will be consistent in the gender distributions of their recommendations.
ALS on BX-E did not personalize at all, so we omit it from analysis.

Table~\ref{tbl:rec-stats} provides the mean tendency for recommendation lists produced by each of our algorithms, plus the tendency of Most Popular and Highest Average Rating recommenders.
Figures \ref{fig:rec-densities-imp} and \ref{fig:rec-densities-exp} show the density of observed recommendation list proportions.

All recommenders were more consistent in their tendencies than the underlying user profiles.

\subsection{From Profiles to Recommendations}
\label{sec:regression}

Our extended Bayesian model (Section~\ref{sec:model-reclists}) allows us to address RQ4: the extent to which our algorithms propagate individual users' tendencies into their recommendations (RQ4).

Figures \ref{fig:rec-densities-imp}--\ref{fig:rec-densities-exp} show the posterior predictive and observed densities of recommender author-gender tendencies, and Figures \ref{fig:rec-regressions-imp}--\ref{fig:rec-regressions-exp} show scatter plots of observed recommendation proportions against user profile proportions with regression curves (regression lines in log-odds space projected into probability space).
Figures \ref{fig:reg-contour-imp}--\ref{fig:reg-contour-exp} show contour plots of the slope and intercept parameters, with lines marking slope of 1 and intercept of 0 (perfect propagation).

In implicit-feedback mode, most algorithms are quite responsive to user profile balances, with slopes greater than 0.5.
Explicit-feedback mode shows less responsiveness and stronger skews: all slopes are relatively small, and intercepts are negative (meaning a user with an evenly-balanced input profile will receive recommendations that have more men than women).
\section{Forced-Balance Recommendation}
\label{sec:aa}
\label{sec:force}

So far we have sought to measure, without intervention, the distribution of author genders of books recommended to users.
This approach is quite reasonable given that neither past work, nor the analysis presented here, is sufficient to inform what recommendations \emph{should} look like.
Individual recommender systems professionals may, through other data, analysis, or philosophy, come to a conclusion about how they want their recommendation algorithms to behave.

In this section we address RQ5 with a suite of \emph{forced-balance recommenders}, that attempt to constrain the distribution in recommender output without substantially impacting recommendation quality.
We consider very simple algorithms for understanding this tradeoff; the behavior of more sophisticated approaches such as calibration \citep{steck_calibrated_2018} or independence \citep{Kamishima2018-ri} are left for future work.
As there is no general definition of "best tradeoff" between quality and gender distribution, nor clear consensus about exactly what to target in the first place, such an analysis would be premature.
Instead we seek to provide lower-limits to what can be expected from these type of tradeoffs with simple approaches.
This analysis serves as a starting point for future explorations into recommender systems that deliberately pursue targeted changes in recommendation properties.

We consider three force-balance recommenders:
\begin{itemize}
    \item single-pass force-balance (SingleEQ)
    \item multi-pass force-balance (GreedyEQ)
    \item multi-pass calibrate (GreedyReflect)
\end{itemize}

All three algorithms are implemented as a  post-processor that can be applied to any recommendation technique, much like \citeauthor{Ziegler2005-zo}'s topic diversification \citep{Ziegler2005-zo}.
This means they operate on the output of another recommendation strategy, re-ranking the items.

The first two algorithms simply attempt to recommend approximately the same number of male- and female-authored books.
In SingleEQ this is accomplished in a single pass of the recommendation list.
This algorithm by rejecting any recommendation that would make an imbalance in the current output list worse (e.g. if current output has more female-authored than male-authored books, it discards female-authored recommendations until the list is balanced)\footnote{This does not accommodate authors with non-binary gender identities. Our goal here is examine the behavior of simple mechanisms supported by available data.}. Note that books with unknown or unlinked gender are always recommended if reached as they cannot increase the gender imbalance (having no listed gender).

The multi-pass algorithm (GreedyEQ) is structured more like a traditional greedy optimizer.
It seeks to optimize the original rank of recommended items subject a gender balance constraint.
This algorithm selects the top-ranked book that is not currently in the output list and that would not make the gender imbalance worse.
The major difference between this and the single pass algorithm is that, while it is slower, it is expected to return better items when runs of books by the same gender are presented by the underlying recommender.
As before, unkown and unlnked authors are recommended as they are reached by this algorithm.

The third and final reranker, multi-pass calibrate (GreedyReflect), is based on Steck's concept of calibration \citep{steck_calibrated_2018}.
Instead of seeking to hit a specific target gender ratio for every user in the system, this algorithm seeks to have as close as possible match between the gender distribution in the user's rating profile and the recommendation list.
The operation of this algorithm is identical to GreedyEQ: at each step it selects the top-ranked book that is not currently recommended so long as it doesn't lead the set of recommended book's gender balance to stray away from our target.
This general framework could easily be adjusted to create any particular target relationship between profile and recommended output gender bias.

We repeated our evaluation from Section~\ref{sec:perf} with the reranking algorithms to measure their accuracy loss.\footnote{We discovered late in preparation that the reranking algorithms were run with an older version of the book-gender file, with about 2\% lower coverage.  We do not expect this significantly impacts accuracy results.}

\begin{figure}[tb]
\includegraphics[width=\columnwidth]{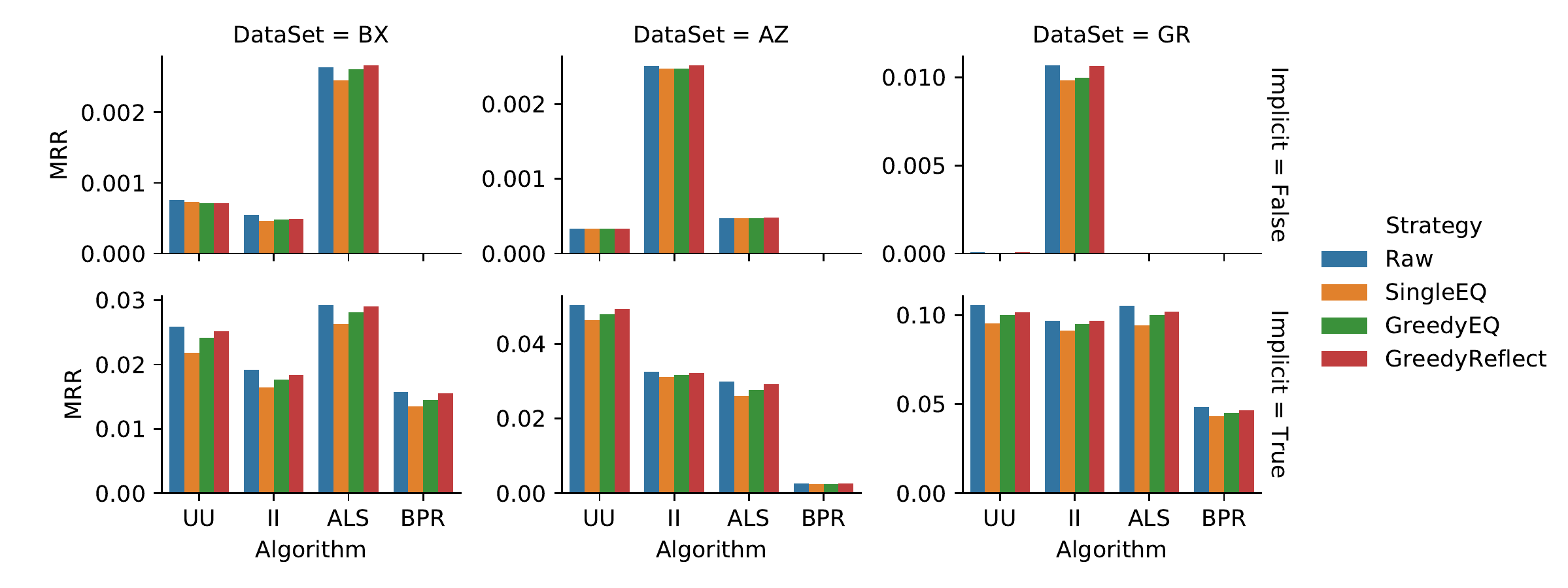}
\caption{Top-$N$ accuracy of natural recommenders and the Forced Balance strategies.}
\label{fig:balance-results}
\end{figure}

Figure~\ref{fig:balance-results} shows the results of this experiment, and Table~\ref{tbl:balance-penalty} shows the relative loss of balancing each algorithm for each data set.  Most penalties are quite small at just a few percent; the largest are (item-item on BX-E, user-user on GR-E) are on algorithms that do not perform well to begin with.
In some cases the balance equalization even improves the recommender's accuracy somewhat.

\begin{table}[tb]
\caption{Accuracy loss for balancing genders.}
\label{tbl:balance-penalty}
\centering{\small\begin{tabular}{llrrr}
\toprule
   Data & Algorithm &  GreedyEQ &  GreedyReflect &  SingleEQ \\
\midrule
\multirow{3}{*}{AZ-E}  & ALS &     0.39\% &         -0.89\% &     1.07\% \\
         & II &     1.26\% &         -0.35\% &     1.32\% \\
         & UU &     0.00\% &          0.00\% &     0.00\% \\
\multirow{4}{*}{AZ-I}    & ALS &     7.52\% &          2.25\% &    13.01\% \\
         & BPR &     6.71\% &          1.14\% &     7.46\% \\
          & II &     2.92\% &          1.05\% &     4.37\% \\
          & UU &     4.71\% &          1.96\% &     7.74\% \\
 \midrule
\multirow{3}{*}{BX-E} &  ALS &     1.14\% &         -0.94\% &     7.09\% \\
          & II &    11.63\% &         10.14\% &    15.58\% \\
          & UU &     6.01\% &          6.09\% &     3.73\% \\
\multirow{3}{*}{BX-I}     & ALS &     3.84\% &          0.45\% &    10.02\% \\
          & BPR &     7.95\% &          1.28\% &    14.40\% \\
          & II &     7.77\% &          4.19\% &    14.10\% \\
          & UU &     6.35\% &          2.52\% &    15.41\% \\
\midrule
\multirow{2}{*}{GR-E} & II &     6.48\% &          0.23\% &     7.87\% \\
          & UU &    22.98\% &         -4.81\% &    43.89\% \\
\multirow{4}{*}{GR-I}   & ALS &     4.79\% &          3.05\% &    10.37\% \\
          & BPR &     7.08\% &          4.18\% &    11.02\% \\
          & II &     1.71\% &         -0.12\% &     5.58\% \\
          & UU &     5.18\% &          3.68\% &     9.61\% \\
\bottomrule
\end{tabular}}
\end{table}

As expected, the multi-pass GreedyEQ algorithm generally outperforms SingleEQ. GreedyReflect, matching the user's profile balance instead of an arbitrary target of 0.5, usually performs the best.

We find, therefore, that it is possible to adjust the recommendation output balance with very simple approaches without substantial loss in accuracy.


\section{Summary of Findings}

To summarize our findings:

\begin{description}
\item[RQ1 --- Baseline Gender Distribution]
Known books are significantly more likely to be written by men than by women; representation among rated books is more balanced.

\item[RQ2 --- User Input Gender Distributions]
User are diffuse in their rating tendencies, with an overall trend favoring male authors but less strongly than the baseline distribution.

\item[RQ3 --- Recommender Output Distributions]
Different CF techniques produce recommendations with quite different distributions.  Matrix factorization on BookCrossing produced reliably male-biased recommendations, while nearest-neighbor and PF techniques were closer to the user profile tendency while being less diffuse than their inputs. Some algorithm and data set combinations resulted in recommendations that were more balanced than their inputs.

\item[RQ4 --- Distribution Propagation]
Implicit-feedback collaborative filters reliably propagated users' input balances into the recommendation outputs.
Explicit-feedback algorithms did not propagate as much.

\item[RQ5 --- Adjusting Balance]
We find that recommendation balance can be adjusted with minimal impact on accuracy.
\end{description}

\section{Discussion}

We found that users in the BookCrossing data set exhibit mild, diffuse tendency towards books written by men; users in the Amazon data set exhibit a somewhat stronger but still highly diffuse tendency.
Both tendencies are more evenly balanced than the set of available of books.
Collaborative filtering algorithms trained on this data exhibit remarkably different behavior; some learn substantially stronger and more consistent tendencies, in some cases producing lists more imbalanced than the item universe.
Others propagate users' tendencies into their recommendation lists.

Nearest-neighbor recommenders in implicit feedback mode also propagated much of each user's profile tendencies into their recommendations.
One interpretation of this is that they are partially picking up on a user's preference for books by male or female authors and reflecting this preference in the recommendations, which is what we would expect from a personalized recommender algorithm. 
The matrix factorization technique we tested consistently exhibited a much stronger bias towards male authors than was present in the input data, and was largely oblivious to individual users' preferences or biases (or, indeed, their book preferences).
It also did not produce very accurate recommendations in our parameter tuning compared to the other algorithms.

The answer to the question ``how do recommenders interact with gender distributions?'' is therefore not simple.  
It has good company with other questions of the social impact of recommendations; for example, contrary to the filter bubble hypothesis, recommender algorithms had a \emph{diversifying} effect on users' viewing portfolios in one movie recommendation service \citep{Nguyen2014-se}.
Exact answers likely depend on algorithm, application, and a number of other variables.

\subsection{Limitations of Data and Methods}
\label{sec:limitations}
Our data and approach has a number of limitations that are important to note.
First, book rating data is extremely sparse, and the BookCrossing data set is small, providing a limited picture of users' reading histories and reducing the performance of some algorithms.
In particular, the high sparsity of the data set caused the MF algorithm to perform particularly poorly on offline accuracy metrics, so these findings may not be representative of its behavior in the wild; future work will need to test them across a range of recommender effectiveness levels and stages of system cold-start.

Second, our data and statistical methods only account for binary gender identities.  While the MARC21 Authority Format supports flexible gender identity records (including multiple possibly-overlapping identities over the course of an author's life and nonbinary identities from an open vocabulary), VIAF does not seem to use this flexibility.

Third, we test a limited set of collaborative filtering algorithms.
While we have chosen algorithms with an eye for diverse behaviors and global popularity, we must acknowledge that our selection of 5 algorithms is small in the face of algorithm diversity in the field.
While our ultimate goal is to understand general trends, we acknowledge that our study does not evaluate enough algorithms to make claims about the entire field.

We consider it valuable to make forward progress in understanding the interaction of information systems with social concerns using the data we have available, even if that data has significant known weaknesses.
We must, however, be reflective and forthright about the limitations of the data, methods, and resulting findings, and seek to improve them in order to develop a better understanding of the human impact of computing systems.  
Our experimental design can be readily extended to accommodate richer or higher-quality data sources and additional algorithms, and the code we provide for our experiments will facilitate such improvements.
We have tested this reproducibility by re-running the experiments in the course of writing and revising this paper.
Ultimately we see this as the first step in untangling a broader issue; we are actively exploring many extensions and improvements to this work.
\section{Conclusion and The Road Ahead}

We have conducted an initial inquiry into the response of collaborative filtering book recommenders to gender distributions in the user preference data on which they are trained. The algorithms differed in their response to these distributions.

This paper is a first step in a much larger project to understand the ways in which recommendation algorithms interact with potentially discriminatory biases, and general behavior of recommendation technology with respect to various social issues. There are many future steps we see for advancing this agenda:

\begin{itemize}
\item Obtaining higher-quality data for measuring distributions of interest in recommender inputs and outputs.
This includes obtaining data on non-binary gender identities and extending our statistical methods to account for them.
\item Examining other content creator features, such as ethnicity, in recommendation applications.
\item Studying other domains and applications, such as movies, research literature, and social media.
\item Develop more advanced algorithms that interact with various user or item characteristics of social concern; these could be developed to reflect organizational or societal goals or to help users further their individual goals \citep{Ekstrand2016-wt}.
\item Study the effect of existing refinements, such as diversification \citep{Willemsen2011-ej,Ziegler2005-zo}, on recommendation distributions.
\end{itemize}

We hope to see more work in the coming years to better understand ways in which recommender systems respond to and influence their sociotechnical contexts.

\section*{Acknowledgements}
This work was partially supported by NSF grant IIS 17-51278.
Mucun Tian, Mohammed R. Imran Kazi, and Hoda Mehrpouyan were authors on the original
conference paper on which this work builds, but did not contribute to the work presented
here.

\bibliographystyle{abbrvnat}      
\bibliography{bag2}   

\end{document}